\documentclass[pra,12pt,tightenlines]{revtex4}
\usepackage{latexsym}
\usepackage{graphicx}
\usepackage{amsmath}
\usepackage{amsfonts}
\usepackage{amsbsy}
\usepackage{amsthm}
\usepackage{bbm}
\usepackage{bm}
\usepackage{dsfont} 
\newtheorem{theorem}{Theorem}

\newtheorem{lemma}{Lemma}

\newcommand{\A}{\mathcal{A}}
\newcommand{\B}{\mathcal{B}}

\newcommand{\E}{\mathcal{E}}

\begin{document}

\title{Entanglement or separability: The choice of how to factorize the algebra of a density matrix}
\author{Walter Thirring} \email{walter.thirring@univie.ac.at}
\affiliation{University of Vienna, Faculty of Physics, Boltzmanngasse 5, A-1090 Vienna,
Austria}
\author{Reinhold A. Bertlmann} \email{reinhold.bertlmann@univie.ac.at}
\affiliation{University of Vienna, Faculty of Physics, Boltzmanngasse 5, A-1090 Vienna,
Austria}
\author{Philipp K\"ohler} \email{philipp.martin.koehler@univie.ac.at}
\affiliation{University of Vienna, Faculty of Physics, Boltzmanngasse 5, A-1090 Vienna,
Austria}
\author{Heide Narnhofer} \email{heide.narnhofer@univie.ac.at}
\affiliation{University of Vienna, Faculty of Physics, Boltzmanngasse 5, A-1090 Vienna,
Austria}


\begin{abstract}

Quantum entanglement has become a resource for the fascinating developments in quantum information and quantum communication during the last decades. It quantifies a certain nonclassical correlation property of a density matrix representing the quantum state of a composite system.
We discuss the concept of how entanglement changes with respect to different factorizations of the algebra which describes the total quantum system. Depending on the considered factorization a quantum state appears either entangled or separable. For pure states we always can switch unitarily between separability and entanglement, however, for mixed states a minimal amount of mixedness is needed. We discuss our general statements in detail for the familiar case of qubits, the GHZ states, Werner states and Gisin states, emphasizing their geometric features. As theorists we use and play with this free choice of factorization, which for an experimentalist is often naturally fixed. For theorists it offers an extension of the interpretations and is adequate to generalizations, as we point out in the examples of quantum teleportation and entanglement swapping.

\end{abstract}

\maketitle


\section{Introduction}\label{sec:introduction}

The surprising features of entanglement in the measurement correlations of two subsystems were highlighted already in 1935 by Einstein, Podolsky and Rosen (EPR) \cite{EPR}. They observed that for a suitably chosen global quantum state of the system the possible outcome of a measurement in laboratory A (called Alice nowadays) depends on the definite -- but free-choice -- measurement in laboratory B (called Bob), no matter how far B is located. Since Einstein rejected a ``spooky action at a distance'' he was forced to conclude that quantum mechanics is an incomplete theory.

In the same year Erwin Schr\"odinger in his trilogy ``On the present situation in quantum mechanics'' \cite{schrodinger1935} considered an EPR-like situation and argued that in quantum mechanics ``the best possible knowledge of a \emph{whole} does not include the best possible knowledge of all its \emph{parts}''. He named such a situation \emph{entanglement}, ``Verschr\"ankung'' in his original Austrian phrasing. This description already comes closest to our modern concept of entanglement ``the \emph{whole} is in a definite (i.e. pure) state, the \emph{parts} taken individually not''.

This discussion about quantum mechanics was dormant for several decades until in 1964 John S. Bell stirred it up again in his celebrated paper ``On the Einstein-Podolsky-Rosen paradox'' \cite{bell1964}, which caused a dramatic change in the quantum mechanical dispute. Bell was able to show that under a strict locality assumption quantum mechanics cannot be completed in the sense of EPR. More precisely, all local realistic theories must satisfy a so-called Bell inequality, a certain combination of expectation values of combined measurements of Alice and Bob, whereas quantum mechanics violates it. Numerous experimental tests in the years that followed, notably by Clauser and Freedman \cite{clauser-freedman1972, clauser1976}, Aspect and collaborators \cite{aspect-grangier-roger1982, aspect-dalibard-roger1982} and Zeilinger and collaborators \cite{weihs-zeilinger1998}, show clearly the violation of a Bell inequality and the confirmation of the quantum mechanical prediction (see, e.g. Ref.~\cite{bertlmann-zeilinger02}).

In the nineties the interests shifted towards quantum information, quantum communication and quantum computation (see, e.g. Ref.~\cite{bowmeester-zeilinger2000}). There the basic ingredient for the quantum states is entanglement, it acts as a resource to allow for certain operations which are otherwise classically impossible to do. One of the most fascinating example is \emph{quantum teleportation} \cite{BBCJPW}, where the properties of an (even unknown) incoming quantum state at Alice's laboratory can be transferred to an outgoing state in Bob's laboratory with help of an EPR pair \cite{bouwmeester-pan-zeilinger-etal, ursin-zeilinger-etal}.

This brings us already to the subject of our Article, the free choice of how to factorize the algebra of a density matrix implying either entanglement or separability of the quantum state. Only with respect to such a factorization it makes sense to talk about entanglement or separability. Quantum teleportation precisely relies on this fact that we can think of different factorizations in which entanglement is localized respectively measurements take place, as we shall discuss in detail in Chapt.~\ref{sec:factorization in physical examples}. Thus we have to focus more closely on entanglement, the magic ingredient of quantum information theory \cite{bertlmann-zeilinger02}. But entanglement with respect to what? The entanglement of quantum states, which are represented by density matrices, is defined with respect to a tensor product structure in Hilbert-Schmidt space. These are tensor products of an algebra of operators or observables.

However, for a given quantum state, it is our freedom of how to factorize the algebra to which a density matrix refers. Thus we may choose! Via global unitary transformations we can switch from one factorization to the other, where in one factorization the quantum state appears entangled, however, in the other not. Consequently, entanglement or separability of a quantum state depends on our choice of factorizing the algebra of the corresponding density matrix, where this choice is suggested either by the set-up of the experiment or by the convenience for the theoretical discussion. This is our basic message.

Considering equivalently to the algebra of the density matrix the tensor product structure of quantum states we find a close connection to the work of Zanardi and collaborators \cite{zanardi-lidar-lloyd2004} who found the same ``democracy between the different tensor product structures, ... , without further physical assumptions no partition has an ontologically superior status with respect to any other'' \cite{zanardi2001}. Thus it's only the interaction, which we consider to determine the density matrix, or the measurement set-up, which fixes the factorization.

For pure states the status is quite clear. Any state can be factorized such that it appears separable up to being maximally entangled depending on the factorization. This fact has been demonstrated already in Ref.~\cite{harshman-ranade}. For mixed states, however, the situation is much more complex (see, e.g., Ref.~\cite{lewenstein-Bruss-etal}). The reason is that the maximal mixed state, the tracial state  $\frac{1}{D}\mathds{1}_D\,$, is separable for any factorization and therefore a sufficiently small neighborhood of it is separable too. Thus the question is how mixed can a quantum state be in order to find a factorization that makes the state as entangled as possible. For a generally mixed state we don't know a precise answer, however, in special cases we do.

In this Article we investigate such special cases for mixed density matrices subjected to certain constraints. In Chapt.~\ref{sec:factorization-algebra} we present our general statements on mixed density matrices and the constraints that make it possible to choose a factorization such that a quantum state appears entangled. In Chapt.~\ref{sec:illustration with qubits} we illustrate our general theorems within the most familiar case of qubits, emphasizing the nice geometric features. We discuss the GHZ states, the Werner states and the Gisin states. The latter ones we particularly present in detail to stress the difference between the local filtering operations, which increase the nonlocal structure of a quantum state and are experimentally feasible, and our unitary transformations which switch between separability and entanglement of a state.

The physical implication of the free choice of factorizing the algebra of a density matrix we discuss in physical examples such as quantum teleportation and entanglement swapping, shedding more light on these amazing quantum phenomena (Chapt.~\ref{sec:factorization in physical examples}). Finally some further conclusions and possible further applications are drawn in Chapt.~\ref{sec:conclusion}.

\section{Factorization algebra}\label{sec:factorization-algebra}

We work in a Hilbert-Schmidt space ${\widetilde{\cal H}}_1 \otimes \widetilde{{\cal H}}_2$ of operators on the finite dimensional bipartite Hilbert space ${\cal H}_1 \otimes {\cal H}_2$, with dimension $D = d_1 \times d_2$. The quantum states $\rho$ (i.e. density matrices) are elements of ${\widetilde{\cal H}}_1 \otimes \widetilde{{\cal H}}_2$ with the properties $\rho^\dag = \rho$, Tr $\rho = 1$ and $\rho \geq 0$. A scalar product on ${\widetilde{\cal H}}_1 \otimes \widetilde{{\cal H}}_2$ is defined by $\left\langle A|B \right\rangle = \textnormal{Tr}\, A^\dag B$ with $A,B \in {{\widetilde{\cal H}}_1 \otimes \widetilde{{\cal H}}_2}$ and the corresponding squared norm is $\|A\|^2=\rm{Tr}\,A^\dag A$.

We consider states over $M^D\,$ corresponding to a density matrix $\rho\,$. Such a state is called separable with respect to the factorization $M^{d_1} \otimes M^{d_2}\,$, if $\rho = \sum_i \rho_{i1} \otimes \rho_{i2}\,$, otherwise it is entangled. Choosing an other factorization $U (M^{d_1} \otimes \mathds{1}) \,U^\dag\,$ and $U (\mathds{1} \otimes M^{d_2}) \,U^\dag\,$, where $U$ represents a unitary transformation on the total space, a former separable state can appear entangled and vice versa. Instead we can consider for a separable $\rho$ the effect of $U$ on $\rho\,$, i.e. $\rho_U = U\rho\,U^\dag\,$, and $\rho_U$ can become entangled for $M^{d_1} \otimes M^{d_2}\,$. This corresponds to the equivalence whether we work in the Schr\"odinger picture or in the Heisenberg picture in the characterization of the quantum states.

Let us first concentrate on pure states, i.e. $\rho\,=\,\left|\,\psi\,\right\rangle\left\langle\,\psi\,\right|\,$, here we prove the following theorem.

\begin{theorem}[Factorization algebra]\ \
For any pure state $\rho$ one can find a factorization $M^D=\A_1 \otimes \A_2$ such that $\rho$ is separable with respect to this factorization and an other factorization $M^D=\B_1 \otimes \B_2$ where $\rho$ appears to be maximally entangled.
\label{theorem:factorization-algebra pure states}
\end{theorem}

\noindent\textbf{Proof:}\ \ \
For each pair of vectors of the same length there are unitary transformations which transform one vector into the other. The vector $|\psi\rangle$ defining the density matrix $\rho\,=\,\left|\,\psi\,\right\rangle\left\langle\,\psi\,\right|\,$ for a pure state can be transformed into any product vector $|\psi_1\rangle \otimes |\psi_2 \rangle\,$ by a unitary operator $U\in M^D\,$, i.e. $U|\psi\rangle = |\psi_1\rangle \otimes |\psi_2 \rangle\,$, or on the other hand we may choose $U$ such that the state is maximally entangled $U|\psi\rangle = \frac{1}{\sqrt{d}} \sum_i |\psi_{1,i}\rangle \otimes |\psi_{2,i}\rangle\,$, where $d = \rm{min}(d_1,d_2)\,$. For the density matrix it means the following. Assuming the density matrix $\rho_{\rm{ent}}$ is entangled within the factorization algebra $M^{d_1} \otimes M^{d_2}\,$ then $\exists \,U\,$: $U \rho_{\rm{ent}} U^\dag = \rho_{\rm{sep}}\,$, i.e. after a unitary transformation the density matrix becomes separable within this factorization. However, transforming also the factorization algebra $M^D = U (M^{d_1} \otimes M^{d_2}) \,U^\dag\,$ we may consider $\rho_{\rm{sep}}\,$ within this unitarily transformed factorization, there it is entangled. Of course, we may choose either factorization $M^D = U (M^{d_1} \otimes M^{d_2}) \,U^\dag\,$ or $M^{d_1} \otimes M^{d_2}\,$ (since $U$ preserves all algebraic relations used in the definitions). \hspace{0.3cm}\mbox{q.e.d.}\\

The extension to mixed states requires some restrictions, as seen from the tracial state $\frac{1}{D}\mathds{1}_D$ which is separable for any factorization.

\begin{theorem}[Factorization in mixed states]\ \
For any mixed state $\rho$ one can find a factorization $M^D=\A_1 \otimes \A_2$ such that $\rho$ is separable with respect to this factorization. An other factorization $M^D=\B_1 \otimes \B_2$ where $\rho$ appears to be entangled exists only beyond a certain bound of mixedness.
\label{theorem:factorization-algebra mixed states}
\end{theorem}

\noindent\textbf{Proof:}\ \ \
Starting with a factorization $M^D=\A_1 \otimes \A_2$ we can find an orthonormal basis (ONB) of separable pure states, namely
$|\varphi_i \rangle \otimes |\psi_j \rangle$ with $i=1,...,d_1,\;\; j=1,...,d_2$ and the set $\{\varphi_i\}$ denotes an ONB in ${\cal H}_1$ and $\{\psi_j\}$ an ONB in ${\cal H}_2\,$. Then every density matrix
\begin{equation}\label{mixed-rho-decomposition}
\rho \;=\; \sum\limits_{\alpha = 1}^{\rm{D}} \,\rho_\alpha \left|\,\chi_{\,\alpha}\,\right\rangle\left\langle\,\chi_{\,\alpha}\,\right| \,,
\end{equation}
where we identify the indices $\{\alpha , \alpha = 1, ... , D\}$ with the set $\{(i,j), i = 1, ... ,d_1, j = 1, ... ,d_2 \}$, can be unitarily transformed into a separable state by $\,U|\,\chi_{\,\alpha}\,\rangle = |\varphi_{i,\alpha} \rangle \otimes |\psi_{j,\alpha} \rangle\,$, i.e.
\begin{equation}\label{separable mixed-rho-transformation}
U \rho \,U^\dag \;=\; \sum\limits_{\alpha = 1}^{\rm{D}} \,\rho_\alpha \left|\,\varphi_{i,\alpha}\,\right\rangle\left\langle\,\varphi_{i,\alpha}\,\right| \otimes
\left|\,\psi_{j,\alpha}\,\right\rangle\left\langle\,\psi_{j,\alpha}\,\right|
\end{equation}
is definitely separable.

On the other hand, we can also find an ONB (see Refs.~\cite{vollbrecht-werner00, werner01, narnhofer06, bertlmann-krammer-AnnPhys09}) of maximally entangled states, where we have chosen $d_1 = d_2 = d$ for simplicity,
\begin{equation}\label{Weyl-basis}
|\,\chi_{kl}\,\rangle \;=\; \sum\limits_{j} \,e^{\frac{2\pi i}{d}jl} \,|\varphi_{j} \rangle \otimes |\psi_{j+k} \rangle \,,
\end{equation}
and a map $U$: $U|\,\chi_{\alpha}\,\rangle = |\,\chi_{kl}\,\rangle\,$, where we again identify the indices $\alpha \leftrightarrow (k,l)$ .

With this unitary transformation our initial density matrix $\rho$ can be turned into a so-called Weyl state
\begin{equation}\label{mixed-rho-transformation to Weyl state}
U \rho \,U^\dag \;=\; \rho_{\,\rm{Weyl}}\,,
\end{equation}
which is expanded into the ONB of maximally entangled states (\ref{Weyl-basis}). However, since the set of entangled states is not convex $\rho_{\,\rm{Weyl}}$ is not automatically entangled.
Note, that state (\ref{mixed-rho-transformation to Weyl state}) is an analogous construction of a Wigner function as demonstrated in Ref.~\cite{narnhofer06}.

For Weyl states a fairly good characterization of the regions of separable states (where the Wigner function remains positive), entangled and bound entangled states exists (see e.g., Refs.~\cite{baumgartner-hiesmayr-narnhofer06, baumgartner-hiesmayr-narnhofer07, baumgartner-hiesmayr-narnhofer08, bertlmann-krammer-JPA08, bertlmann-krammer-AnnPhys09, horodecki99, bertlmann-krammer-PRA-78-08, bertlmann-krammer-PRA-77-08}). \hspace{0.3cm}\mbox{q.e.d.}\\

In $2 \times 2$ dimensions the constraints of the set $\{\rho_\alpha\}$ in order to characterize the regions of separability and entanglement are also well-known \cite{verstraete-audenaert-demoor}. Especially for the Werner states, see Sec.~\ref{sec:Werner states}, one immediately sees that the choice (\ref{Weyl-basis}) need not be optimal. In fact, it can be quite unfavorable as we show in the example below, see Sec.~\ref{sec:theorem Kus-Zyczkowski}, here a decomposition into one separable state and remaining entangled states is optimal. \\

\noindent\textbf{Remark:}\ \ \
For $d_1 = d_2 = 2\,$ the set of separable states among the Weyl states is known due to the Peres--Horodecki criterion~\cite{peres96, horodecki96}, a criterion that is necessary for separability in any dimensions $d_1 \times d_2$ and sufficient for $2 \times 2$ and $2 \times 3$ dimensional Hilbert spaces. Accordingly, a separable state has to stay positive semidefinite under partial transposition (PT), it is called a PPT state. Thus, if a density matrix becomes indefinite under PT, i.e. one or more eigenvalues are negative, it has to be entangled and we call it a NPT state.\\

Clearly, we also want to formulate Theorem~\ref{theorem:factorization-algebra mixed states} more precisely. Let us consider a generalized Werner state in $d \times d$ dimensions (thus we choose $d_1 = d_2 = d$)
\begin{equation} \label{werner state decomposition into projector}
    \rho \;=\; \alpha P \,+\, \frac{1-\alpha}{d^2}\,\mathds{1}_{d^2}\qquad \mbox{with} \;
    \;\; 0 \leq \alpha \leq 1 \,,
\end{equation}
where $P$ is a projector ($P^2 = P$) to a maximally entangled state. The maximal eigenvalue of $\rho$ (\ref{werner state decomposition into projector}) is $\alpha + \frac{1}{d^2}\,$. Then we find the following lemma.
\begin{lemma}[Bound for splitted states]\ \
Assume a state can be split into a maximally entangled state, corresponding to a projector $P$, and an orthogonal state $\sigma$
\begin{equation} \label{werner state decomposition into orthogonal states}
    \rho \;=\; \beta P \,+\, (1-\beta)\,\sigma \qquad \mbox{with} \;
    \left\langle P |\sigma \right\rangle \;=\; 0 \quad \mbox{and} \;\; 0 \leq \beta \leq 1 \,.
\end{equation}
Then the following statement holds: \ \ If $\beta > \frac{1}{d}$ the state (\ref{werner state decomposition into orthogonal states}) is entangled.
\label{lemma:bound for werner state}
\end{lemma}

\noindent Note, for $d=2$ it implies the well-known bound $\alpha > \frac{1}{3}$ for the Werner states, see Sec.~\ref{sec:Werner states}, and in matrix form decomposition (\ref{werner state decomposition into orthogonal states}) can be written as
\begin{equation}
\rho \;=\;
\begin{pmatrix}
\alpha + \frac{1-\alpha}{d^2} & 0                    & .        & .     & 0 \\
0                             & \frac{1-\alpha}{d^2} &          &       & \\
.                             &                      & .        &       & \\
.                             &                      &          & .     & \\
0                             &                      &          &       & \frac{1-\alpha}{d^2} \\
\end{pmatrix}\,.
\label{Werner-state into orthogonal states in matrix notation}
\end{equation}

\noindent\textbf{Proof:}\ \ \
We prove Lemma~\ref{lemma:bound for werner state} by using the following optimal entanglement witness (explained more explicitly in Sec.~\ref{sec:Alice-Bob})
\begin{equation}\label{entanglement witness for Werner states}
A \;=\; \mathds{1}_{d^2} \,-\, d \,P
\end{equation}
to find a bound of separability or entanglement. The inner product of a witness with all separable states remains positive semidefinite
\begin{equation}\label{werner entanglement witness and separable states}
\left\langle \rho_{\rm{sep}} | A \right\rangle \;=\; \textnormal{Tr}\, \rho_{\rm{sep}} A \;\geq\; 0 \,.
\end{equation}
Thus we have to show that the expectation value of the entanglement witness is positive semidefinite for all separable states, i.e. for all product states
\begin{eqnarray} \label{Werner EW for separable states}
    & &\left\langle \varphi \otimes \psi | \,\mathds{1}_{d^2} \,-\, d \,P\, | \varphi \otimes \psi \right\rangle \;=\; 1 \,-\, d\,\left\langle \varphi \otimes \psi | P | \varphi \otimes \psi \right\rangle \nonumber\\
    & &=\; 1 \,-\, d\,\sum_{i,j =1}^d \frac{1}{d}\,\varphi_i^\ast\psi_i^\ast\varphi_j\psi_j
    \;=\; 1 \,-\, \left\langle \varphi^\ast | \psi \right\rangle \left\langle \psi | \varphi^\ast \right\rangle
    =\; 1 \,-\, \mid\left\langle \varphi^\ast | \psi \right\rangle\mid^2 \;\geq\; 0 \,,
\end{eqnarray}
since $\mid\left\langle \varphi^\ast | \psi \right\rangle\mid \leq 1\,$, and $\left\langle \varphi \otimes \psi | \,\mathds{1}_{d^2} \,-\, d \,P\, | \varphi \otimes \psi \right\rangle \;=\; 0 \;\; \mbox{iff} \;\; | \psi \rangle \,=\, | \varphi^\ast \rangle \,$, which makes the witness optimal.

Applying now the entanglement witness (\ref{entanglement witness for Werner states}) to the state (\ref{werner state decomposition into orthogonal states}) we get for entanglement
\begin{eqnarray} \label{bound for beta}
0 \;>\; \left\langle \,\beta P \,+\, (1-\beta)\,\sigma \,|\, \mathds{1}_{d^2} \,-\, d \,P \,\right\rangle \;=\; 1 \,-\, \beta \,d
\quad \Rightarrow \quad \beta \;>\; \frac{1}{d} \,. \hspace{0.3cm}\mbox{q.e.d.}
\end{eqnarray}

Note, the bound is optimal for Werner states (see Sec.~\ref{sec:Werner states}) and the states of the Gisin line (see Sec.~\ref{sec:Gisin states}). However, if not all eigenvalues of $\sigma$ are equal then the witness $A \,=\, \mathds{1}_{d^2} - d \,P\,$ is not optimal with respect to $\rho$ (\ref{werner state decomposition into orthogonal states}), but in this case the decomposition (\ref{werner state decomposition into orthogonal states}) does not represent any more a Werner state.

Concluding, each state with maximal eigenvalue $\rho_{\rm{max}} > \frac{1}{d}\,$ can be factorized like in Eq.~(\ref{werner state decomposition into orthogonal states}) and is entangled with respect to this factorization. However, generally this factorization is not optimal. Under certain constraints we now find a factorization which is indeed optimal.\\

Let $\rho$ be any mixed state with an ordered spectrum $\{\rho_1 \geq \rho_2 \geq ... \geq \rho_{d^2 - 2} \geq \rho_{d^2 - 1} \geq \rho_{d^2}\}\,$, thus we have the decomposition (again we have chosen $d_1 = d_2 = d$ without loss of generality)
\begin{equation}\label{rho into projectors of eigenstates}
\rho \;=\; \rho_1 P_1 \,+\, \rho_2 P_2 \,+\, ...\,+\, \rho_{d^2} P_{d^2} \,,
\end{equation}
where $P_1, P_2, ... , P_{d^2}\,$ are the projectors to the corresponding eigenstates. Furthermore we consider maximal entanglement in a two-dimensional subspace, specifically we choose
\begin{eqnarray}
P_1 \;\rightarrow\; Q_1 &\;=\;& \frac{\rho_1}{2}\,\left|\,11 + 22\,\right\rangle\left\langle\,11 + 22\,\right| \nonumber\\
P_{d^2 -2} \;\rightarrow\; Q_2 &\;=\;& \rho_{d^2 - 2}\,\left|\,12\,\right\rangle\left\langle\,12\,\right| \nonumber\\
P_{d^2} \;\rightarrow\; Q_3 &\;=\;& \rho_{d^2}\,\left|\,21\,\right\rangle\left\langle\,21\,\right| \nonumber\\
P_{d^2 - 1} \;\rightarrow\; Q_4 &\;=\;& \frac{\rho_{d^2 - 1}}{2}\,\left|\,11 - 22\,\right\rangle\left\langle\,11 - 22\,\right|\,.
\label{eq:projectors in 2-dim subspace}
\end{eqnarray}
Then we find the following theorem.

\begin{theorem}[Factorization under constraints]\ \
If $\rho_1 > \frac{3}{d^2}\,$, i.e. the largest eigenvalue $\rho_1$ is bounded below by $\frac{3}{d^2}$ then there is always a choice of factorization possible such that the partial algebras are entangled.
\label{theorem:factorization-algebra constrained}
\end{theorem}

\noindent\textbf{Proof:}\ \ \
To find entanglement we consider the partially transposed of matrix $\rho$ (\ref{rho into projectors of eigenstates}) with choice (\ref{eq:projectors in 2-dim subspace}), it contains the following structure
\begin{equation}
\rho^{\,\rm{PT}} \;=\;
\begin{pmatrix}
. & .                                & .                                & . \\
. &	\rho_{d^2 - 2}                   & \frac{1}{2}(\rho_1 - \rho_{d^2 - 1}) & . \\
. &	\frac{1}{2}(\rho_1 - \rho_{d^2 - 1}) & \rho_{d^2}                   & . \\
. & .                                & .                                & . \\
\end{pmatrix}\,.
\label{eq:rho PT matrix structure}
\end{equation}
Due to the Peres--Horodecki criterion~\cite{peres96, horodecki96} it is entangled, i.e. a NPT state, if it contains a negative eigenvalue. This is the case if $\rho_{d^2 - 2}\cdot\rho_{d^2} \,-\, \frac{1}{4}(\rho_1 - \rho_{d^2 - 1})^2 < 0$, which is a geometric mean
\begin{equation}\label{geometric mean of eigenvalues}
\sqrt{\rho_{d^2 - 2}\cdot\rho_{d^2}} \,<\, \frac{1}{2}(\rho_1 - \rho_{d^2 - 1})\,.
\end{equation}
We relax the estimate by replacing the geometric mean by the arithmetic mean
\begin{eqnarray} \label{arithmetic mean of eigenvalues}
& &\frac{\rho_{d^2 - 2} \,+\, \rho_{d^2}}{2} \;<\; \frac{1}{2}(\rho_1 - \rho_{d^2 - 1}) \nonumber\\
& &\Rightarrow \quad \rho_{d^2 - 2} \,+\, \rho_{d^2 - 1} \,+\, \rho_{d^2} \;<\; \rho_1 \,.
\end{eqnarray}
On the other hand we have
\begin{equation}
\rho_{d^2 - 2} \,+\, \rho_{d^2 - 1} \,+\, \rho_{d^2} \;<\; \frac{1 \,-\, \rho_1 }{d^2 \,-\, 3} \,,
\end{equation}
leading to
\begin{equation}
3\,\frac{1 \,-\, \rho_1 }{d^2 \,-\, 3} \;<\; \rho_1 \qquad \mbox{or} \qquad \rho_1 \;>\; \frac{3}{d^2} \,. \hspace{0.4cm}\mbox{q.e.d.}
\end{equation}

\noindent Note, these arguments are similar to those leading to Lemma~\ref{lemma:absolute separability in 2x2 dimensions}.\\

Thus, under the constraints of Theorem~\ref{theorem:factorization-algebra constrained} a mixed state is separable with respect to some factorization and entangled with respect to another.

It is interesting now to search for those states which are separable with respect to all possible factorizations of the composite system into subsystems $\A_1 \otimes \A_2\,$. This is the case if $\rho_{\rm{U}} \,=\, U\rho\,U^\dag$ remains separable for any unitary transformation $U$. Such states are called \emph{absolutely separable states} \cite{kus-zyczkowski, zyczkowski-bengtsson, bengtsson-zyczkowski-book}, the tracial state being the prototype. In this connection the \emph{maximal ball} of states around the tracial state $\frac{1}{d^2}\,\mathds{1}_{d^2}$ with a general radius $r = \frac{1}{d^2 - 1}$ of constant mixedness is considered, which can be inscribed into the separable states (see Refs.~\cite{kus-zyczkowski, gurvits-barnum}). This radius is given in terms of the Hilbert-Schmidt distance
\begin{equation}\label{radius by HS distance}
d\,(\rho , \mathds{1}_{d^2}) \;=\; \left\| \rho \,-\, \frac{1}{d^2}\,\mathds{1}_{d^2} \right\| \;=\; \sqrt{\rm{Tr}\,\big(\rho \,-\, \frac{1}{d^2}\,\mathds{1}_{d^2}\big)^2}\,.
\end{equation}
Notice that in the different topologies the relevant parameters scale in the same way with the dimension $d$.

\begin{theorem}[Absolute separability of the Ku\'s-\.Zyczkowski ball \cite{kus-zyczkowski}]\ \
All states belonging to the maximal ball which can be inscribed into the set of mixed states for a bipartite system are not only separable but also absolutely separable.
\label{theorem:Kus-Zyczkowski ball}
\end{theorem}

Note that geometrically in case of Theorem~\ref{theorem:factorization-algebra constrained} the set of absolutely separable states is not as symmetric as in case of Theorem~\ref{theorem:Kus-Zyczkowski ball}, the set is even a bit larger containing the maximal ball and corresponds rather to a ``Laberl'' \footnote{``Laberl'' or more precise ``Fetzenlaberl'' is the Viennese expression for a self-made football out of shreds, which in the old times the Viennese boys liked to play with on the streets.} than to a ball.

Furthermore, we already know that in the two-qubit case the set of absolutely separable states is larger than the maximal ball of  Ku\'s and \.Zyczkowski. As conjectured in Ref.~\cite{ishizaka-hiroshima} and proved in Ref.~\cite{verstraete-audenaert-demoor} the set of absolutely separable states contains any mixed state with certain constraints on the spectrum.
\begin{lemma}[Absolute separability in $2 \times 2$ dimensions \cite{verstraete-audenaert-demoor}]\ \
Let $\rho$ be any mixed state in $2 \times 2$ dimensions with an ordered spectrum $\{\rho_1 \geq \rho_2 \geq \rho_3 \geq \rho_4\}\,$.

\noindent If the spectrum is constrained by the inequality $\rho_1 \,-\, \rho_3 \,-\, 2\sqrt{\rho_2\rho_4} \;\leq\; 0 \,,$ then $\rho$ is absolutely separable.
\label{lemma:absolute separability in 2x2 dimensions}
\end{lemma}

As an example we want to quote the state with spectrum $\{0.47,0.30,0.13,0.10 \}$ (see Ref.~\cite{kus-zyczkowski}) that does not belong to the maximal ball but satisfies the constraints of Lemma~\ref{lemma:absolute separability in 2x2 dimensions} and is, for this reason, absolutely separable.

\section{Illustration with qubits}\label{sec:illustration with qubits}

\subsection{Geometry of physical states}\label{sec:geometry of physical states}

\begin{figure}
	\centering
	\includegraphics[width=0.55\textwidth]{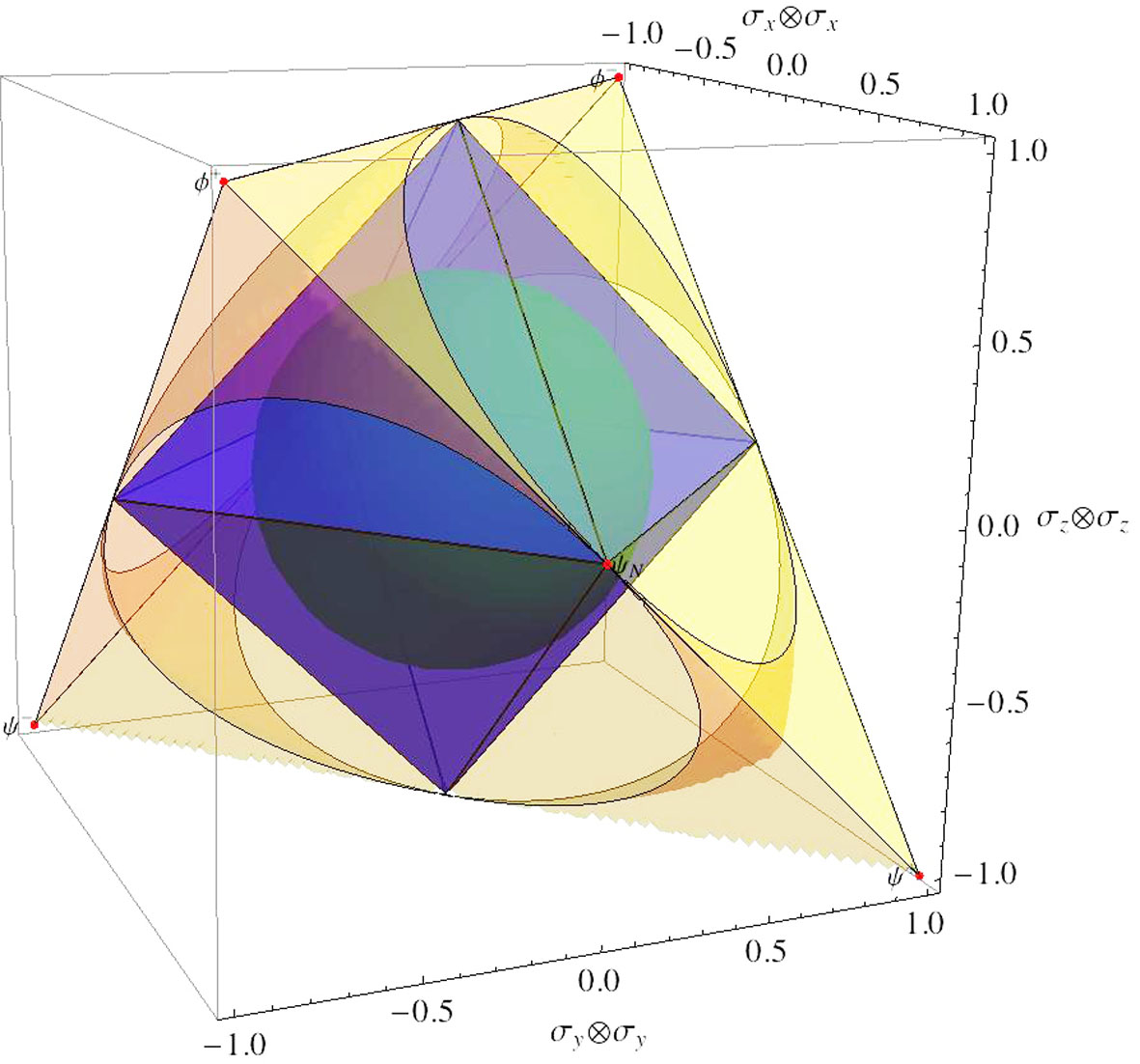}
	\caption{Tetrahedron of physical states in $2 \times 2$ dimensions spanned by the four Bell states $\psi^+ , \psi^- , \phi^+ , \phi^-\,$: The separable states form the blue double pyramid and the entangled states are located in the remaining tetrahedron cones. The unitary invariant Ku\'s-\.Zyczkowski ball (shaded in green) is placed within the double pyramid and the maximal mixture $\frac{1}{4}\mathds{1}_4$ is at the origin. Outside the ball at the corner of the double pyramid is the state $\psi_{\rm{N}}$ (\ref{eq:narnhofer state in matrix notation}), the separable state with maximal purity. The local states according to a Bell inequality lie within the dark-yellow surfaces containing all separable but also some entangled states.}
	\label{fig:tetrahedron of physical states}
\end{figure}

Geometrically all Weyl states in $2 \times 2$ dimensions, the celebrated case of Alice and Bob in quantum information, lie within a tetrahedron (three-dimensional simplex) spanned by the four maximally entangled Bell states $\left|\,\psi^{\,\pm}\,\right\rangle , \left|\,\phi^{\,\pm}\,\right\rangle\,$, it is the domain of the physical states, see Fig.\ref{fig:tetrahedron of physical states}.
The separable states (convex set) form a double pyramid (shaded in blue) within the tetrahedron. The entangled states are located in the tetrahedron cones outside of the double pyramid and in the middle (at the origin) rests the maximal mixed, the tracial state $\frac{1}{4}\mathds{1}_4$ (see Refs.~\cite{bertlmann-narnhofer-thirring02, vollbrecht-werner-PRA00, horodecki-R-M96}).

The set of local states, satisfying a Bell inequality \`a la CHSH, defines a domain (shaded by the dark-yellow surfaces) that is, interestingly, much larger than the area of separable states (see also Ref.~\cite{spengler-huber-hiesmayr09}).

Within the double pyramid the Ku\'s-\.Zyczkowski ball of absolutely separable states \cite{kus-zyczkowski} (shaded in green) is placed, whose radius of constant mixedness is determined by the nearest separable state to a Bell state. All states within this maximal ball remain separable for any unitary transformation (Theorem~\ref{theorem:Kus-Zyczkowski ball}).

\subsection{Illustration of Theorem~\ref{theorem:Kus-Zyczkowski ball}}\label{sec:theorem Kus-Zyczkowski}

To illustrate Theorem~\ref{theorem:Kus-Zyczkowski ball} the following example is quite instructive. Let us choose a separable state outside of the maximal ball, say at a corner of the double pyramid, see Fig.~\ref{fig:tetrahedron of physical states}.
\begin{figure}
	\centering
	\includegraphics[width=0.55\textwidth]{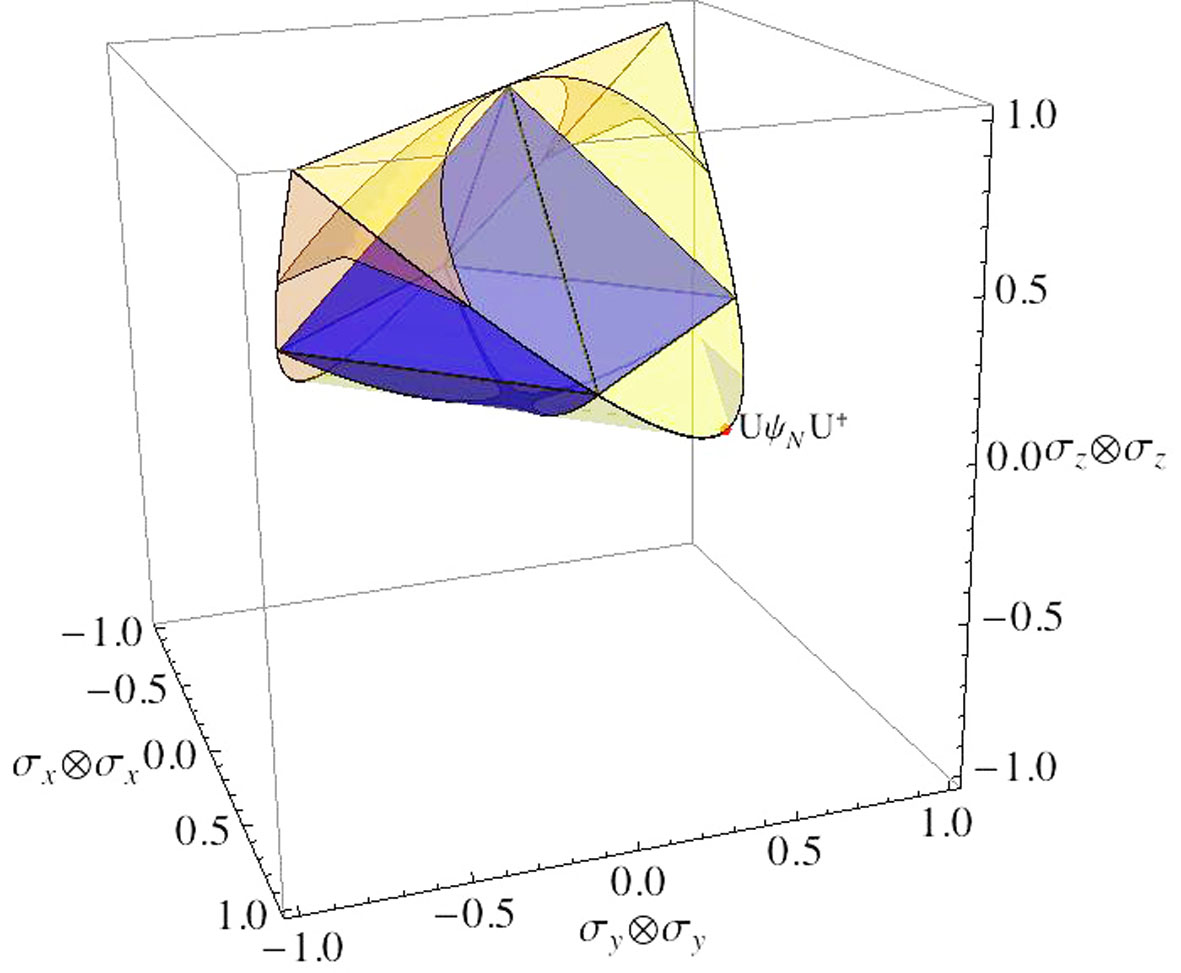}
	\caption{Terms like ($\sigma_z\,\otimes\,\mathds{1} \,+\, \mathds{1}\,\otimes\,\sigma_z$) in the density matrix affect the region of physical states, the entangled, local and separable areas shrink. In particular, the former tetrahedron of Weyl states becomes parabolic in z-direction such that the unitarily transformed state $U\psi_{\rm{N}}\,U^\dag$ (\ref{eq:narnhofer state unitary transformed}) slips into the entangled domain (yellow part), being a maximally entangled mixed state (MEMS). At origin the state $\frac{1}{4} \big( \,\mathds{1}\,\otimes\,\mathds{1} \,+\, \frac{1}{2}\,(
\sigma_z\,\otimes\,\mathds{1} \,+\, \mathds{1}\,\otimes\,\sigma_z\,)\big)$ is located.}
	\label{fig:tetrahedron shrinkage of physical states}
\end{figure}
It is given by the matrix
\begin{eqnarray}
\rho_{\rm{N}} \;=\; \left|\,\psi_{\rm{N}}\,\right\rangle\left\langle\,\psi_{\rm{N}}\,\right| \;=\; \frac{1}{2}\,(\rho^+ \,+\, \omega^+) \;=\; \frac{1}{4}
\begin{pmatrix}
1  & 0 & 0 & 1 \\
0  & 1 & 1 & 0 \\
0  & 1 & 1  & 0 \\
1  & 0 & 0  & 1 \\
\end{pmatrix}\,,
\label{eq:narnhofer state in matrix notation}
\end{eqnarray}
or in Bloch decomposition
\begin{equation}\label{eq:narnhofer state in Bloch decomposition}
\rho_{\rm{N}} \;=\; \frac{1}{4}\left(\mathds{1} \otimes \mathds{1} \,+\, \sigma_x \otimes \sigma_x\right)\,,
\end{equation}
and this separable state has the smallest possible mixedness or largest purity. It is the purity $P(\rho) = \textnormal{Tr}\,\rho^2$ that is a way to quantify the degree of mixedness, especially adjusted for the geometry of a state $\rho$ and it ranges between $\frac{1}{d} \leq P(\rho) \leq 1\,$.

The following unitary transformation
\begin{eqnarray}
U &\;=\;& \frac{1}{\sqrt{2}}\,
\begin{pmatrix}
 1 & 0        & 0        & 1 \\
 0 & \sqrt{2} & 0        & 0 \\
 0 & 0        & \sqrt{2} & 0 \\
-1 & 0        & 0        & 1 \\
\end{pmatrix}\nonumber\\
&\;=\;& \frac{1}{4} \big( (2+\sqrt{2})\,\mathds{1}\,\otimes\,\mathds{1} \,+\, i\sqrt{2}\,(
\sigma_x\,\otimes\,\sigma_y \,+\, \sigma_y\,\otimes\,\sigma_x\,)
\,-\, (2-\sqrt{2})\,\sigma_z\,\otimes\,\sigma_z \,\big)
\label{eq:unitary transform for narnhofer state}
\end{eqnarray}
transforms the state $\rho_{\rm{N}} = \left|\,\psi_{\rm{N}}\,\right\rangle\left\langle\,\psi_{\rm{N}}\,\right|$ into
\begin{eqnarray}
\rho_{\rm{U}} &\;=\;& U\rho_{\rm{N}}\,U^\dag \;=\; \frac{1}{4}
\begin{pmatrix}
2 & 0 & 0 & 0 \\
0 & 1 & 1 & 0 \\
0 & 1 & 1 & 0 \\
0 & 0 & 0 & 0 \\
\end{pmatrix}\nonumber\\
&\;=\;& \frac{1}{4} \big( \,\mathds{1}\,\otimes\,\mathds{1} \,+\, \frac{1}{2}\,(
\sigma_z\,\otimes\,\mathds{1} \,+\, \mathds{1}\,\otimes\,\sigma_z\,)
\,+\, \frac{1}{2}\,(\sigma_x\,\otimes\,\sigma_x \,+\, \sigma_y\,\otimes\,\sigma_y )\,\big) \,.
\label{eq:narnhofer state unitary transformed}
\end{eqnarray}
However, due to the occurrence of the term ($\sigma_z\,\otimes\,\mathds{1} \,+\, \mathds{1}\,\otimes\,\sigma_z$) the transformation $U$ (\ref{eq:unitary transform for narnhofer state}) leads to a quantum state outside of the set of Weyl states. This new state $\rho_{\rm{U}}$ (\ref{eq:narnhofer state unitary transformed}) is not positive any more under partial transposition, $\rho_{\rm{U}}^{\rm{PT}} \not\geq 0\,$, where $\rho_{\rm{U}}^{\rm{PT}} = (\mathds{1}\,\otimes\,T_{\rm{B}} )\,\rho_{\rm{U}}$
and $T_{\rm{B}}$ means partial transposition on Bob's subspace. Therefore, due to the Peres-Horodecki criterion the state $\rho_{\rm{U}}$ (\ref{eq:narnhofer state unitary transformed}) is entangled with the concurrence $C = \frac{1}{2}\,$. Transformation $U$ (\ref{eq:unitary transform for narnhofer state}) is already optimal, i.e. it entangles $\rho_{\rm{N}}$ maximally. Thus $\rho_{\rm{U}}$ belongs to the so-called MEMS class, the class of maximally entangled mixed states for a given value of purity \cite{ishizaka-hiroshima, munro-james-white-kwiat}.

In order to illustrate states of type $\rho_{\rm{U}}\,$, i.e. states with additional terms ($\sigma_z\,\otimes\,\mathds{1} \,+\, \mathds{1}\,\otimes\,\sigma_z$) we see that these additional degrees of freedom affect the region of physical states (where $\rho > 0)\,$, see Fig.~\ref{fig:tetrahedron shrinkage of physical states}. The areas of entangled, local and separable states shrink, in particular, the former tetrahedron of Weyl states becomes parabolic in the lower z-direction such that the unitarily transformed state $\rho_{\rm{U}} = U\rho_{\rm{N}}\,U^\dag$ (\ref{eq:narnhofer state unitary transformed}) slips into the entangled domain (yellow part in Fig.~\ref{fig:tetrahedron shrinkage of physical states}), and is maximally entangled. At origin the state $\frac{1}{4} \big( \,\mathds{1}\,\otimes\,\mathds{1} \,+\, \frac{1}{2}\,(\sigma_z\,\otimes\,\mathds{1} \,+\, \mathds{1}\,\otimes\,\sigma_z\,)\big)$ is located.

\subsection{Alice and Bob}\label{sec:Alice-Bob}

Let us begin with the example of two qubits, the case of Alice and Bob. Here the dimensions of the submatrices are $d_1=d_2=2$ and we span the two $M^2$ factors with aid of two sets of Pauli matrices $\vec \sigma_A$ and $\vec \sigma_B$, the subalgebras of Alice and Bob. The standard product basis is $|\uparrow \rangle \otimes |\uparrow \rangle$, $|\uparrow \rangle \otimes |\downarrow \rangle$, $|\downarrow \rangle \otimes |\uparrow \rangle$, $|\downarrow \rangle \otimes |\downarrow \rangle\,$ and the four maximally entangled Bell vectors are given by
$\left|\,\psi^{\,\pm}\,\right\rangle = \frac{1}{\sqrt{2}}\,\left(\,
		\left|\,\uparrow\,\right\rangle\,\left|\,\downarrow\,\right\rangle\,\pm\,
		\left|\,\downarrow\,\right\rangle\,\left|\,\uparrow\,\right\rangle\,\right)$ and
$\left|\,\phi^{\,\pm}\,\right\rangle = \frac{1}{\sqrt{2}}\,\left(\,
		\left|\,\uparrow\,\right\rangle\,\left|\,\uparrow\,\right\rangle\,\pm\,
		\left|\,\downarrow\,\right\rangle\,\left|\,\downarrow\,\right\rangle\,\right)\,$.
Considering the corresponding density matrices we have for the separable product states explicitly
\begin{equation}
\rho_{\uparrow\uparrow}\,=\,
\begin{pmatrix}
1 & 0 & 0 & 0 \\
0 &	0 & 0 &	0 \\
0 &	0 & 0 &	0 \\
0 & 0 & 0 & 0 \\
\end{pmatrix}\,,\;
\rho_{\uparrow\downarrow}\,=\,
\begin{pmatrix}
0 & 0 & 0 & 0 \\
0 &	1 & 0 &	0 \\
0 &	0 & 0 &	0 \\
0 & 0 & 0 & 0 \\
\end{pmatrix}\,,
\rho_{\downarrow\uparrow}\,=\,
\begin{pmatrix}
0 & 0 & 0 & 0 \\
0 &	0 & 0 &	0 \\
0 &	0 & 1 &	0 \\
0 & 0 & 0 & 0 \\
\end{pmatrix}\,,\;
\rho_{\downarrow\downarrow}\,=\,
\begin{pmatrix}
0 & 0 & 0 & 0 \\
0 &	0 & 0 &	0 \\
0 &	0 & 0 &	0 \\
0 & 0 & 0 & 1 \\
\end{pmatrix}\,,
\label{eq:rho upup-updown-etc in matrix notation}
\end{equation}
and for the entangled Bell states
\begin{equation}
\rho^{\mp} \,=\, \left|\,\psi^\mp\,\right\rangle\left\langle\,\psi^\mp\,\right| \,=\, \frac{1}{2}
\begin{pmatrix}
0 & 0     & 0     & 0 \\
0 &	1     & \mp 1 &	0 \\
0 &	\mp 1 & 1     &	0 \\
0 & 0     & 0     & 0 \\
\end{pmatrix}\,,\;
\omega^{\mp} \,=\, \left|\,\phi^\mp\,\right\rangle\left\langle\,\phi^\mp\,\right| \,=\, \frac{1}{2}
\begin{pmatrix}
1     & 0 & 0 & \mp 1 \\
0     &	0 & 0 &	0 \\
0     & 0 & 0 &	0 \\
\mp 1 & 0 & 0 & 1 \\
\end{pmatrix}\,.
\label{eq:rho Bell-states in matrix notation}
\end{equation}
The unitary matrix $U$ which transforms the entangled basis into the separable one is following
(we suppress from now on the labels A and B for the subspaces)
\begin{equation}
U \,=\, \frac{1}{\sqrt{2}}\left(\mathds{1} \otimes \mathds{1} + i\,\sigma_x \otimes \sigma_y\right) \,=\, \frac{1}{\sqrt{2}}
\begin{pmatrix}
1  & 0 &  0 & 1 \\
0  & 1 & -1 & 0 \\
0  & 1 & 1  & 0 \\
-1 & 0 & 0  & 1 \\
\end{pmatrix}\,.
\label{eq:unitary-transform in matrix notation}
\end{equation}
It is more illustrative to work with the Bloch decompositions of the states to show the subalgebras explicitly. A general state $\rho$ can be decomposed into (see, e.g. Refs.~\cite{bertlmann-narnhofer-thirring02, bertlmann-krammer-JPA08})
\begin{equation} \label{rho-general Bloch decomposition}
\rho \,=\, \frac{1}{4}\left(\,\mathds{1}\,\otimes\,\mathds{1} + r_i \,\sigma_i\,\otimes\,\mathds{1} +
u_i \,\mathds{1}\,\otimes\,\sigma_i + t_{ij} \;\sigma_i\,\otimes\,\sigma_j\right) \,,
\end{equation}
and for separable states we have
\begin{equation} \label{rho-separable Bloch decomposition}
\rho_{\rm{sep}} \,=\, \frac{1}{4}\left(\,\mathds{1}\,\otimes\,\mathds{1} + r_i \,\sigma_i\,\otimes\,\mathds{1} +
u_i \,\mathds{1}\,\otimes\,\sigma_i + r_i u_j \;\sigma_i\,\otimes\,\sigma_j\right) \,,
\end{equation}
with ${\vec r}^{\,2} = {\vec u}^{\,2} = 1$. In particular we find \cite{bertlmann-narnhofer-thirring02}
\begin{eqnarray} \label{rho-Bell-minus Bloch decomposition}
\rho^- \,&=&\,\frac{1}{4}\left(\mathds{1}\,\otimes\,\mathds{1} \,-\, \vec\sigma\,\otimes\,\vec\sigma\right)
\label{rho-updown Bloch decomposition} \\
\rho_{\uparrow\downarrow}\,&=&\,\frac{1}{4}\left(\mathds{1}\,\otimes\,\mathds{1} \,+\, \sigma_z\,\otimes\,\mathds{1} \,-\, \mathds{1}\,\otimes\,\sigma_z \,-\, \sigma_z\,\otimes\,\sigma_z\right) \,.
\end{eqnarray}
Then $U$ transforms the two sets of algebras of Alice and Bob as follows:
\begin{eqnarray}
\sigma_x\,\otimes\,\mathds{1} &\xrightarrow{U}& \sigma_x\,\otimes\,\mathds{1} \qquad\qquad\quad\,
\mathds{1}\,\otimes\,\sigma_x \xrightarrow{U} \sigma_x\,\otimes\,\sigma_z \label{eq:U-transform x-comp} \\
			[2mm]
\sigma_y\,\otimes\,\mathds{1} &\rightarrow& -\sigma_z\,\otimes\,\sigma_y \qquad\qquad\,
\mathds{1}\,\otimes\,\sigma_y \rightarrow \mathds{1}\,\otimes\,\sigma_y \label{eq:U-transform y-comp} \\
			[2mm]
\sigma_z\,\otimes\,\mathds{1} &\rightarrow& \sigma_y\,\otimes\,\sigma_y \qquad\qquad\quad
\mathds{1}\,\otimes\,\sigma_z \rightarrow -\sigma_x\,\otimes\,\sigma_x \,, \label{eq:U-transform z-comp}
\end{eqnarray}
and implies a change in the Alice--Bob tensor products as
\begin{eqnarray}
\sigma_x\,\otimes\,\sigma_x &\xrightarrow{U}& \mathds{1}\,\otimes\,\sigma_z \label{eq:sigma-sigma transform x-comp} \\
			[2mm]
\sigma_y\,\otimes\,\sigma_y &\rightarrow& -\sigma_z\,\otimes\,\mathds{1} \label{eq:sigma-sigma transform y-comp} \\
			[2mm]
\sigma_z\,\otimes\,\sigma_z &\rightarrow& \sigma_z\,\otimes\,\sigma_z \,. \label{eq:sigma-sigma transform z-comp}
\end{eqnarray}

Now, let's consider entanglement. Quite generally, entanglement can be ``detected'' by an Hermitian operator, the so-called \emph{entanglement witness} $A$, that detects the entanglement of a state $\rho_{\rm ent}$ via the \emph{entanglement witness inequalities} (EWI) \cite{horodecki96, terhal00, bruss02, bertlmann-narnhofer-thirring02}
\begin{eqnarray} \label{def-entwit}
    \left\langle \rho_{\rm ent}|A \right\rangle \;=\; \textnormal{Tr}\, \rho_{\rm ent} A
    & \;<\; & 0 \,,\nonumber\\
    \left\langle \rho|A \right\rangle = \textnormal{Tr}\, \rho A & \;\geq\; & 0 \qquad
    \forall \rho \in S \,,
\end{eqnarray}
where $S$ denotes the set of all separable states. An entanglement witness is ``optimal'', denoted by $A_{\rm{opt}}\,$, if apart from Eq.~(\ref{def-entwit}) there exists a separable state $\rho_0 \in S$ such that
\begin{equation}
    \left\langle \rho_0 |A_{\rm{opt}} \right\rangle \;=\; 0 \,.
\end{equation}
The operator $A_{\rm{opt}}$ defines a tangent plane to the convex set of
separable states $S$ and can be constructed in the following way \cite{bertlmann-narnhofer-thirring02}:
\begin{equation} \label{entwitmaxviolation}
    A_{\rm{opt}} \;=\;
    \frac{\rho_0 - \rho_{\rm ent} \,-\, \left\langle \rho_0 ,
    \rho_0 - \rho_{\rm ent} \right\rangle \mathds{1}}{\left\| \rho_0 - \rho_{\rm ent} \right\|} \;,
\end{equation}
where $\rho_0$ represents the nearest separable state.

In particular, for the optimal entanglement witness of the Bell state $\rho^-$ we get
\begin{equation} \label{entwit rho-Bell minus}
A_{\rm{opt}}^{\rho^-} \;=\; \frac{1}{2\sqrt{3}}\left(\mathds{1}\,\otimes\,\mathds{1} \,+\, \vec\sigma\,\otimes\,\vec\sigma\right)\,,
\end{equation}
leading to the EWI
\begin{eqnarray} \label{EWI rho-Bell-minus}
    \left\langle \rho^-|A_{\rm{opt}}^{\rho^-} \right\rangle &\;=\;& \textnormal{Tr}\, \rho^- A_{\rm{opt}}^{\rho^-}
    \,=\, -\frac{1}{\sqrt{3}} \;<\; 0 \,,\nonumber\\
    \left\langle \rho_{\rm{sep}}|A_{\rm{opt}}^{\rho^-} \right\rangle &\;=\;& \textnormal{Tr}\, \rho_{\rm{sep}} A_{\rm{opt}}^{\rho^-} \,=\, \frac{1}{2\sqrt{3}}(1 + \cos \delta) \;\geq\; 0 \qquad
    \forall \rho \in S \,,
\end{eqnarray}
where $\delta$ represents the angle between the unit vectors $\vec r$ and $\vec u$.\\

Transforming now the entangled Bell state $\rho^-\,$ according to Eq.~(\ref{eq:unitary-transform in matrix notation}) we find
\begin{eqnarray} \label{U-transform of rho-Bell minus}
    & & U\,\rho^-\,U^\dag \;=\; \frac{1}{4}\left(\mathds{1}\,\otimes\,\mathds{1} \,+\, \sigma_z\,\otimes\,\mathds{1} \,-\, \mathds{1}\,\otimes\,\sigma_z \,-\, \sigma_z\,\otimes\,\sigma_z\right) \;\equiv\; \rho_{\uparrow\downarrow}\,, \\
    & & \left\langle U\,\rho^-\,U^\dag|A_{\rm{opt}}^{\rho^-} \right\rangle \;=\; \textnormal{Tr}\; U\,\rho^-\,U^\dag \,A_{\rm{opt}}^{\rho^-} \,=\, 0 \;,
\end{eqnarray}
i.e. separability with respect to the algebra $\{ \sigma_i\,\otimes\,\sigma_j \}\,$. Thus the transformed state $U\,\rho^-\,U^\dag$ represents a separable pure state as claimed in Theorem~\ref{theorem:factorization-algebra pure states} and geometrically it has the Hilbert-Schmidt (HS) distance
\begin{equation} \label{HSdistance of rho-Bell minus to transformed state}
    d\,(\rho^-) \;=\; \left\| U\,\rho^-\,U^\dag \,-\, \rho^- \right\| \;=\; 1\,,
\end{equation}
to the state $\rho^-\,$.
This distance represents the amount of entanglement, more precise, it is the Hilbert-Schmidt measure that can be considered as a measure of entanglement and it is defined by \cite{bertlmann-narnhofer-thirring02, bertlmann-krammer-AnnPhys09}
\begin{equation} \label{def-HSmeasure}
    D(\rho_{\rm{ent}}) \;:=\; \min_{\rho \in S} \left\| \rho \,-\, \rho_{\rm{ent}} \right\|
    \;=\; \left\| \rho_0 \,-\, \rho_{\rm{ent}} \right\| \,,
\end{equation}
where $\rho_0$ denotes the nearest separable state, the minimum of the HS distance. In our case of the maximal entangled Bell state $\rho^-$ the nearest separable state is mixed and will be considered in the Section Werner states.

It is interesting that the maximal violation of the EWI (\ref{def-entwit}) for an entangled state is equal to its HS measure (\ref{def-HSmeasure}), the measure of entanglement (Theorem of Ref.~\cite{bertlmann-narnhofer-thirring02}).\\

Transforming on the other hand also the entanglement witness, i.e. choosing a different algebra,
\begin{equation} \label{U-transform of EW-Bell minus}
U\,A_{\rm{opt}}^{\rho^-}\,U^\dag \;=\; \frac{1}{4}\left(\mathds{1}\,\otimes\,\mathds{1} \,-\, \sigma_z\,\otimes\,\mathds{1} \,+\, \mathds{1}\,\otimes\,\sigma_z \,+\, \sigma_z\,\otimes\,\sigma_z\right) \,,
\end{equation}
we then get
\begin{eqnarray} \label{U-transform of EWI}
    \left\langle U \rho^-\,U^\dag|U A_{\rm{opt}}^{\rho^-}\,U^\dag \right\rangle &\;=\;&
    \left\langle \rho^-|A_{\rm{opt}}^{\rho^-} \right\rangle \;=\; -\frac{1}{\sqrt{3}} \;<\; 0 \,,
\end{eqnarray}
and the transformed state is entangled again with respect to the other algebra factorization $\{ \sigma_i\,\otimes\,\mathds{1}, \mathds{1}\,\otimes\,\sigma_j , \sigma_i\,\otimes\,\sigma_j \}\,$. It demonstrates nicely the content of Theorem~\ref{theorem:factorization-algebra pure states} and the analogy of choosing either the Schr\"odinger picture or the Heisenberg picture in the characterization of the quantum states.\\

\begin{figure}
	\centering
	\includegraphics[width=0.55\textwidth]{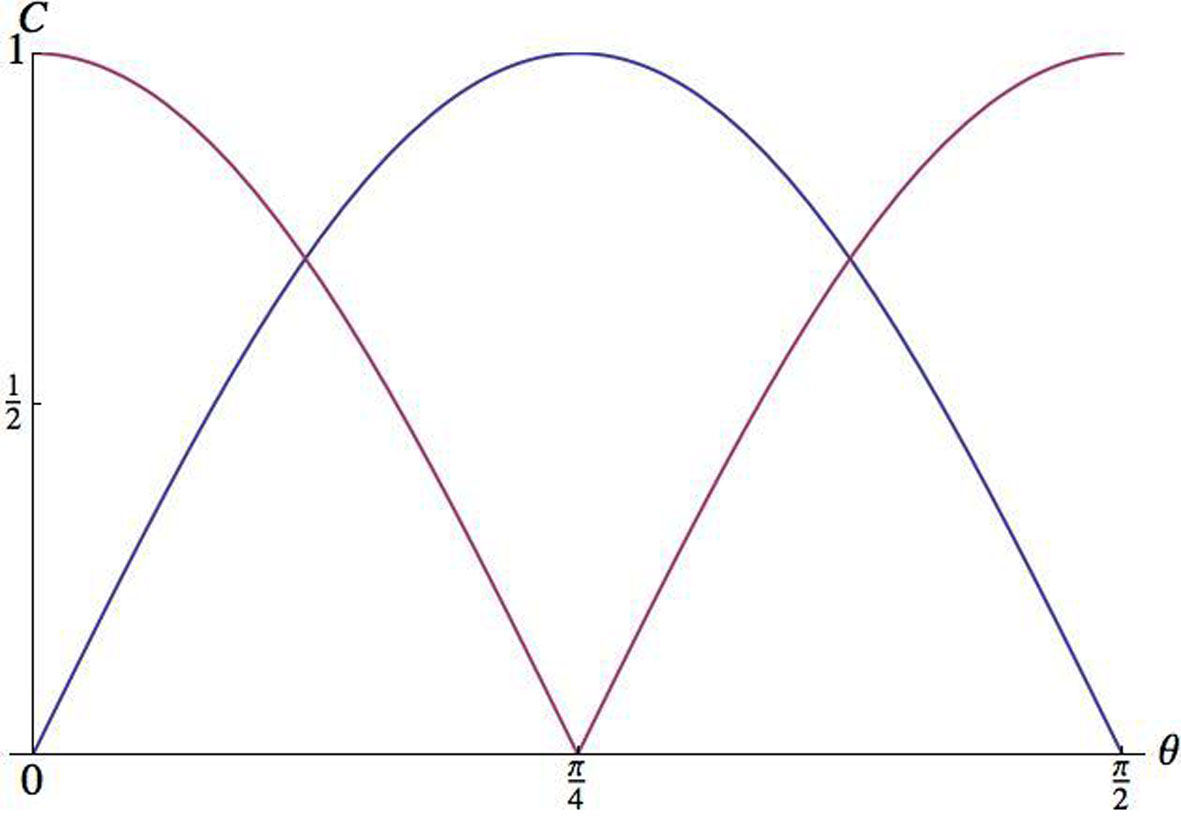}
	\caption{The concurrence of the $\rho_\theta$ state, Eq.~(\ref{rho-theta in Bloch decomposition}) (blue line),
    and the transformed $U \rho_{\,\theta}\,U^\dag$ state, Eq.~(\ref{U-transform of rho-theta in Bloch decomposition}) (red line),  is plotted versus $\theta$.}
	\label{fig:concurrence of theta-state and transformed state}
\end{figure}

Next we study non-maximal entangled states like $\left|\,\psi_{\,\theta}\,\right\rangle = \sin\theta\,
\left|\,\uparrow\,\right\rangle\,\left|\,\downarrow\,\right\rangle \,-\,
\cos\theta\,
\left|\,\downarrow\,\right\rangle\,\left|\,\uparrow\,\right\rangle\,$ with the corresponding density matrix
\begin{equation}
\rho_{\,\theta} \;=\; \left|\,\psi_{\,\theta}\,\right\rangle\left\langle\,\psi_{\,\theta}\,\right| \;=\;
\begin{pmatrix}
0 & 0                          & 0                          & 0 \\
0 &	\sin^2\theta               & -\frac{1}{2}\sin (2\theta) & 0 \\
0 &	-\frac{1}{2}\sin (2\theta) & \cos^2\theta               & 0 \\
0 & 0                          & 0                          & 0 \\
\end{pmatrix}\,,
\label{eq:rho theta-state in matrix notation}
\end{equation}
and the Bloch decomposition
\begin{eqnarray} \label{rho-theta in Bloch decomposition}
\rho_\theta \;=\; & &\frac{1}{4} \big( \,\mathds{1}\,\otimes\,\mathds{1} \,-\,
\cos (2\theta)\,(\sigma_z\,\otimes\,\mathds{1} \,-\, \mathds{1}\,\otimes\,\sigma_z ) \nonumber\\
& &\,-\, \sin (2\theta)\,(\sigma_x\,\otimes\,\sigma_x \,+\, \sigma_y\,\otimes\,\sigma_y)
\,-\, \sigma_z\,\otimes\,\sigma_z \,\big) \,.
\end{eqnarray}
Transforming the state $\rho_\theta\,$ by the unitary transformation (\ref{eq:unitary-transform in matrix notation}) we obtain
\begin{equation}
U\rho_{\,\theta}\,U^\dag \;=\; \frac{1}{2}
\begin{pmatrix}
0 & 0                & 0                & 0 \\
0 &	1+\sin (2\theta) & -\cos (2\theta)  & 0 \\
0 &	-\cos (2\theta)  & 1-\sin (2\theta) & 0 \\
0 & 0                & 0                & 0 \\
\end{pmatrix}\,,
\label{U-transform of rho theta-state in matrix notation}
\end{equation}
and in Bloch form
\begin{eqnarray} \label{U-transform of rho-theta in Bloch decomposition}
U\rho_{\,\theta}\,U^\dag \;=\;  & &\frac{1}{4} \big( \,\mathds{1}\,\otimes\,\mathds{1} \,+\,
\sin (2\theta)\,(\sigma_z\,\otimes\,\mathds{1} \,-\, \mathds{1}\,\otimes\,\sigma_z ) \nonumber\\
& &\,-\, \cos (2\theta)\,(\sigma_x\,\otimes\,\sigma_x \,+\, \sigma_y\,\otimes\,\sigma_y )
\,-\, \sigma_z\,\otimes\,\sigma_z \,\big) \,.
\end{eqnarray}
This transformed state still contains some entanglement (except for $\theta = \frac{\pi}{4}\,$, the Bell state $\rho^-$), which we determine via the concurrence of Wootters \cite{wootters98, hill-wootters97, wootters01} as a measure of entanglement. For the concurrence we first consider the flipped state $\widetilde{\rho}$ of $\rho$
\begin{equation}\label{rho-flipped}
\widetilde{\rho} \;=\; (\sigma_y\,\otimes\,\sigma_y) \,\rho^*\, (\sigma_y\,\otimes\,\sigma_y) \,,
\end{equation}
where $\rho^*$ is the complex conjugate and is taken in the standard product basis and then calculate the concurrence $C$ by the formula
\begin{equation}\label{def-concurrence}
C(\rho) \;=\; \rm{max}\{ 0,\lambda_1 - \lambda_2 - \lambda_3 - \lambda_4 \} \,.
\end{equation}
The $\lambda_i$'s are the square roots of the eigenvalue, in decreasing order, of the matrix $\rho\widetilde{\rho}\,$. As results we find for the state $\rho_{\,\theta}$ and its $U$ transformation the following concurrences
\begin{equation}\label{concurrences rho-theta and U-trans-rho-theta}
C(\rho_{\,\theta}) \;=\; \sin (2\theta) \qquad 
C(U\rho_{\,\theta}\,U^\dag) \;=\; \cos (2\theta) \,,
\end{equation}
which we have plotted in Fig.\ref{fig:concurrence of theta-state and transformed state}.
We see, only for the values $\theta = 0, \frac{\pi}{4}$ the unitary transformation $U$ is the optimal choice to switch between entanglement and separability. For other values of $\theta$ the transformation $U_\theta$ can be adjusted according to
$\,U_\theta \,\left|\,\psi_{\,\theta}\,\right\rangle \;=\; \left|\,\psi^+\,\right\rangle$, providing the following unitary matrix, where $f_{\pm}(\theta) \,=\, \cos\theta \pm \sin\theta\,$,
\begin{eqnarray}\nonumber
U_\theta &\;=\;& \frac{1}{\sqrt{2}}\left(\,f_{-}(\theta)\,\mathds{1} \otimes \mathds{1} \,-\, i\,f_{+}(\theta)\,\sigma_x \otimes \sigma_y\,\right)\\
&\;=\;& \frac{1}{\sqrt{2}}
\begin{pmatrix}
f_{-}(\theta) &  0             & 0             & -f_{+}(\theta) \\
0             &  f_{-}(\theta) & f_{+}(\theta) &  0 \\
0             & -f_{+}(\theta) & f_{-}(\theta) &  0 \\
f_{+}(\theta) &  0             & 0             &  f_{-}(\theta) \\
\end{pmatrix}\,.
\label{eq:unitary-theta-transform in matrix notation}
\end{eqnarray}
Then the such transformed $\rho_{\,\theta}\,$ state is maximal entangled and represents the Bell state $\rho^+$
\begin{equation}\label{unitary-transform of rho-theta}
U_\theta \,\rho_{\,\theta}\,U_\theta^\dag \;=\; \rho^+ \qquad \forall\theta \,.
\end{equation}
Both unitary transformation in succession
\begin{eqnarray}\nonumber
\widetilde{U_\theta} &\;=\;& U\,U_\theta \;=\; \cos\theta \;\mathds{1} \otimes \mathds{1} - i\,\sin\theta \;\sigma_x \otimes \sigma_y\\
&\;=\;&
\begin{pmatrix}
\cos\theta & 0           &  0         & -\sin\theta \\
0          & \cos\theta  & \sin\theta & 0 \\
0          & -\sin\theta & \cos\theta & 0 \\
\sin\theta & 0           & 0          & \cos\theta \\
\end{pmatrix}
\label{eq:combined unitary-theta-transform}
\end{eqnarray}
clearly make $\rho_{\,\theta}$ separable for all $\theta$
\begin{equation}\label{combined unitary-transform of rho-theta}
\widetilde{U_\theta} \,\rho_{\,\theta}\,\widetilde{U_\theta}^\dag \;=\; \rho_{\downarrow\uparrow} \qquad \forall\theta \,,
\end{equation}
which demonstrates the content of Theorem~\ref{theorem:factorization-algebra pure states}.

\subsection{GHZ states}\label{sec:GHZ states}

What we have illustrated in the case of Alice \& Bob we also find in a system of three qubits when tracing over one subspace. This leads us to the popular GHZ states \cite{GHZ89, Pan-etal00} which play an important role in the fundamentals and techniques of quantum information (see, e.g. Ref~\cite{bertlmann-zeilinger02}).

The usual three--photon GHZ state is defined by
\begin{equation}\label{GHZ VH-state }
\left|\,\psi^{\,\rm{GHZ}}\,\right\rangle \;=\; \frac{1}{\sqrt{2}}\,\big(\,
		\left|\,V\,\right\rangle\,\left|\,V\,\right\rangle\,\left|\,V\,\right\rangle \,+\,
		\left|\,H\,\right\rangle\,\left|\,H\,\right\rangle\,\left|\,H\,\right\rangle\,\big)\,,
\end{equation}
where $V$ and $H$ denote vertical and horizontal polarizations, respectively. We slightly generalize the state, as we did before, to non-maximal entanglement and use now the bit notation $|0\rangle$ and $|1\rangle$ of quantum information, then we get
\begin{equation}\label{GHZ theta-state }
\left|\,\psi^{\,\rm{GHZ}}_{\,\theta}\,\right\rangle \;=\; \sin\theta\,
		\left|\,0\,\right\rangle\,\left|\,0\,\right\rangle\,\left|\,0\,\right\rangle \,+\,
\cos\theta\,
		\left|\,1\,\right\rangle\,\left|\,1\,\right\rangle\,\left|\,1\,\right\rangle\,,
\end{equation}
yielding the density matrix
\begin{equation}\label{rho GHZ theta-state}
\rho^{\,\rm{GHZ}}_{\,\theta} \;=\; \left|\,\psi^{\,\rm{GHZ}}_{\,\theta}\,\right\rangle\left\langle\,\psi^{\,\rm{GHZ}}_{\,\theta}\,\right|\,.
\end{equation}
Next we trace over one subsystem -- we don't count the outcome of this subsystem -- and are left with the state of a bipartite system
\begin{equation}\label{rho GHZ theta-state matrix}
\widetilde{\rho}^{\,\,\rm{GHZ}}_{\,\theta} \;=\; \rm{Tr}_{\rm{subsystem}}\, \rho^{\,\rm{GHZ}}_{\,\theta} \;=\;
\begin{pmatrix}
\sin^2\theta & 0 & 0 & 0 \\
0            & 0 & 0 & 0 \\
0            & 0 & 0 & 0 \\
0            & 0 & 0 & \cos^2\theta \\
\end{pmatrix}\,,
\end{equation}
or in the Bloch decomposition this state is expressed by
\begin{eqnarray} \label{rho GHZ theta-state Bloch decomposition}
\widetilde{\rho}^{\,\,\rm{GHZ}}_{\,\theta} \;=\; & &\frac{1}{4} \big( \,\mathds{1}\,\otimes\,\mathds{1} \,-\,
\cos (2\theta)\,(\sigma_z\,\otimes\,\mathds{1} \,+\, \mathds{1}\,\otimes\,\sigma_z)
\,+\, \sigma_z\,\otimes\,\sigma_z \,\big) \,.
\end{eqnarray}
State (\ref{rho GHZ theta-state matrix}), (\ref{rho GHZ theta-state Bloch decomposition}) is a mixed state and separable $\forall \,\theta\,$.
Note, due to the tracing over one subsystem we bring back GHZ to a special case of Alice \& Bob.

Now we find unitary transformations such that they entangle the separable state maximally. In the interval $0 \leq \theta \leq \frac{\pi}{4}$ the unitary transformation
\begin{equation}
U_1 \;=\; \frac{1}{\sqrt{2}}
\begin{pmatrix}
\sqrt{2} & 0  & 0        & 0 \\
0        & 1  & 0        & -1  \\
0        & -1 & 0        & -1 \\
0        & 0  & \sqrt{2} & 0 \\
\end{pmatrix}
\label{eq:U_1-transform of GHZ theta in matrix notation}
\end{equation}
is best. For the remaining part of the interval $\frac{\pi}{4} \leq \theta \leq \frac{\pi}{2}$ we use the transformation
\begin{equation}
U_2 \;=\; \frac{1}{\sqrt{2}}
\begin{pmatrix}
0 & 0        & 0  & \sqrt{2} \\
1 & 0        & -1 & 0 \\
1 & 0        & 1  & 0 \\
0 & \sqrt{2} & 0  & 0 \\
\end{pmatrix}\,.
\label{eq:U_2-transform of GHZ theta in matrix notation}
\end{equation}
The Bloch decompositions of both transformation matrices (\ref{eq:U_1-transform of GHZ theta in matrix notation}) and (\ref{eq:U_2-transform of GHZ theta in matrix notation}) are quite elaborate and will be skipped. Transformations $U_1$ and $U_2$ are already optimal, i.e. lead to maximally entangled states (for a definition of optimal transformations, see Ref.~\cite{verstraete-audenaert-demoor}).

For the such transformed GHZ state we find the following matrix expressions
\begin{equation}
U_1 \,\rho^{\,\,\rm{GHZ}}_{\,\theta}\, U_1^\dag \;=\; \frac{1}{2}\,
\begin{pmatrix}
2\sin^2\theta & 0            & 0            & 0 \\
0             & \cos^2\theta & \cos^2\theta & 0 \\
0             & \cos^2\theta & \cos^2\theta & 0 \\
0             & 0            & 0            & 0 \\
\end{pmatrix}\,,
\label{rho GHZ theta-state U_1 transformed in matrix}
\end{equation}
\begin{equation}
U_2 \,\rho^{\,\,\rm{GHZ}}_{\,\theta}\, U_2^\dag \;=\; \frac{1}{2}\,
\begin{pmatrix}
2\cos^2\theta & 0            & 0            & 0 \\
0             & \sin^2\theta & \sin^2\theta & 0 \\
0             & \sin^2\theta & \sin^2\theta & 0 \\
0             & 0            & 0            & 0 \\
\end{pmatrix}\,,
\label{rho GHZ theta-state U_2 transformed in matrix}
\end{equation}
or in Bloch decompositions we have
\begin{eqnarray}\label{rho GHZ theta-state U_1 transformed in Bloch decomposition}
U_1 \,\rho^{\,\,\rm{GHZ}}_{\,\theta}\, U_1^\dag \;=\; & &\frac{1}{4} \big( \,\mathds{1}\,\otimes\,\mathds{1} \,+\,
\sin^2\theta\,(\sigma_z\,\otimes\,\mathds{1} \,+\, \mathds{1}\,\otimes\,\sigma_z ) \\ \nonumber
& &\,+\, \cos^2\theta \,(\sigma_x\,\otimes\,\sigma_x \,+\, \sigma_y\,\otimes\,\sigma_y )
\,-\, \cos (2\theta)\,\sigma_z\,\otimes\,\sigma_z \,\big) \,,
\end{eqnarray}
\begin{eqnarray}\label{rho GHZ theta-state U_2 transformed in Bloch decomposition}
U_2 \,\rho^{\,\,\rm{GHZ}}_{\,\theta}\, U_2^\dag \;=\; & &\frac{1}{4} \big( \,\mathds{1}\,\otimes\,\mathds{1} \,+\,
\cos^2\theta\,(\sigma_z\,\otimes\,\mathds{1} \,+\, \mathds{1}\,\otimes\,\sigma_z ) \\ \nonumber
& &\,+\, \sin^2\theta \,(\sigma_x\,\otimes\,\sigma_x \,+\, \sigma_y\,\otimes\,\sigma_y )
\,+\, \cos (2\theta)\,\sigma_z\,\otimes\,\sigma_z \,\big) \,.
\end{eqnarray}
These transformed states are symmetric in the exchange of $\cos\theta \longleftrightarrow \sin\theta\,$.

\begin{figure}
	\centering
	\includegraphics[width=0.55\textwidth]{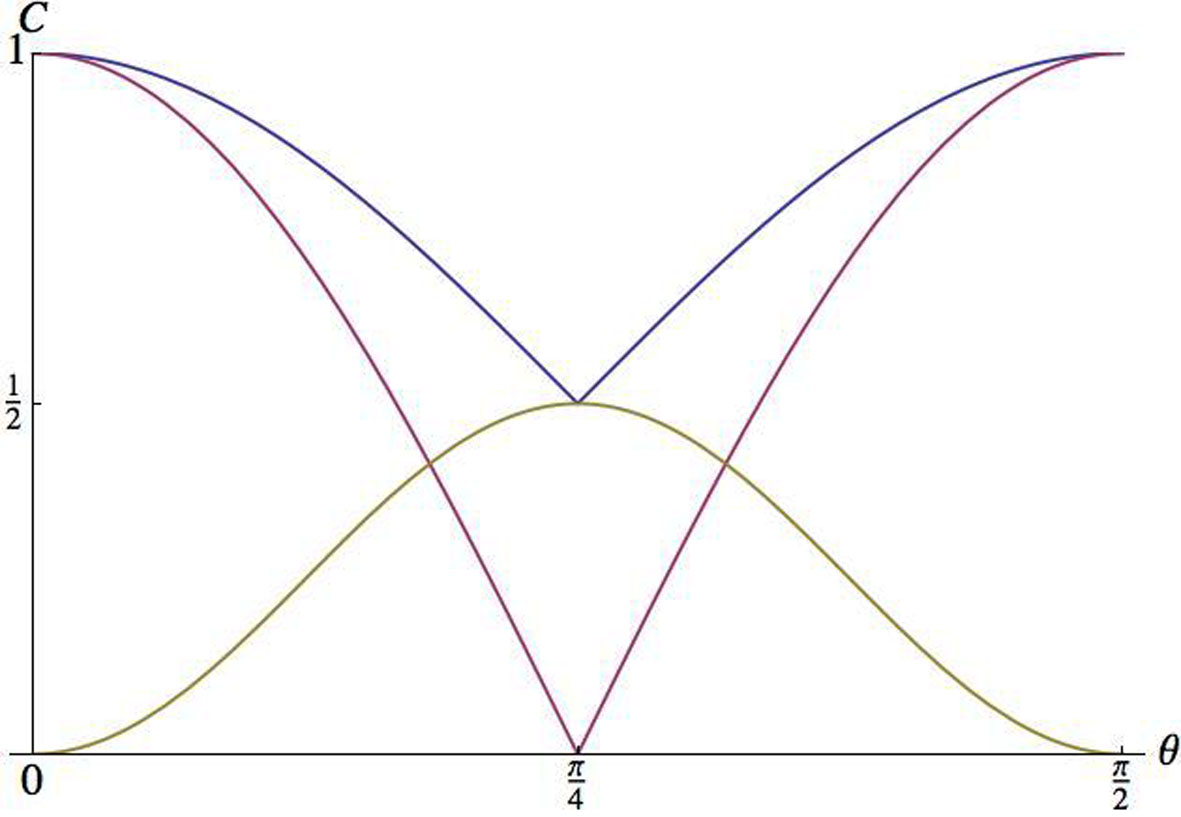}
	\caption{The concurrence of the transformed $U_1 \,\rho^{\,\,\rm{GHZ}}_{\,\theta}\, U_1^\dag$ GHZ state, Eq.~(\ref{rho GHZ theta-state U_1 transformed in Bloch decomposition}), for the interval $0 \leq \theta \leq \frac{\pi}{4}$ and $U_2 \,\rho^{\,\,\rm{GHZ}}_{\,\theta}\, U_2^\dag$, Eq.~(\ref{rho GHZ theta-state U_2 transformed in Bloch decomposition}), for $\frac{\pi}{4} \leq \theta \leq \frac{\pi}{2}$ (blue line), is plotted versus $\theta$. For comparison the effect of the $U$ transformation (\ref{eq:unitary-transform in matrix notation}) on the GHZ state is shown (red line) and its mixedness (yellow line).}
	\label{fig:concurrence of transformed GHZ theta state}
\end{figure}

The amount of entanglement we calculate via the concurrence of the state as we did before in the case of Alice \& Bob and for comparison we also compute the mixedness defined by
\begin{equation}\label{def:mixedness}
\delta \;=\; 1 \,-\, \rm{Tr}\,\big( \widetilde{\rho}^{\,\,\rm{GHZ}}_{\,\theta}\big)^2\,.
\end{equation}
The results we have plotted in Fig.\ref{fig:concurrence of transformed GHZ theta state}.
Whereas the transformation $U$ (\ref{eq:unitary-transform in matrix notation}), working for Alice \& Bob, does not entangle the separable GHZ state (\ref{rho GHZ theta-state matrix}) at the value $\theta = \frac{\pi}{4}\,$, the unitary transformations (\ref{eq:U_1-transform of GHZ theta in matrix notation}) and (\ref{eq:U_2-transform of GHZ theta in matrix notation}) do. Since the transformations (\ref{eq:U_1-transform of GHZ theta in matrix notation}) and (\ref{eq:U_2-transform of GHZ theta in matrix notation}) entangle the separable $\widetilde{\rho}^{\,\,\rm{GHZ}}_{\,\theta}$ maximally, i.e. the transformed states belong to the MEMS class, all possible entangling unitary transformations lie in between the two (blue and red) curves. The maximal value of the concurrence $C = \frac{1}{2}$ at $\theta = \frac{\pi}{4}$ coincides with the one of the mixedness $\delta = \frac{1}{2}$, which is specific for our considered state (\ref{rho GHZ theta-state matrix}) but does not hold in general.\\

These observations on the GHZ state we can generalize by the following lemma.

\begin{lemma}[Entanglement for traced GHZ states]\ \
Let $|\Omega\rangle$ be a pure state on the tensor product of the algebras $\A_1 \otimes \A_2 \otimes \A_3$ with dimensions $d_1, d_2, d_3 = d_1 = d\,$. Let furthermore
\begin{equation}\label{reduced omega matrix of GHZ}
    \big(\left|\,\Omega\,\right\rangle\left\langle\,\Omega\,\right|\big)_{\A_3} \;=\; \rho_{\A_3} \;=\;
\begin{pmatrix}
\rho_1 & 0    & \cdot & \cdot \\
0      &\cdot &       &  \\
\cdot  &      & \cdot &   \\
\cdot  &      &       & \rho_d \\
\end{pmatrix}
\end{equation}
be the reduced density matrix on $\A_3\,$: $\rho_{\A_3} \;=\; \rm{Tr}_{\A_1 \otimes \A_2}\,\left|\,\Omega\,\right\rangle\left\langle\,\Omega\,\right|\,$.

\noindent Then there exists a unitary transformation $U$ in $\A_1 \otimes \A_2$ such that the amount of entanglement $\E(\A_i,\A_3)$ between the subalgebras $\A_i \; (i=1,2)$ and $\A_3$ is given by
\begin{eqnarray}\label{entanglement of subalgebras}
    \E(U\,\A_1 \otimes \mathds{1}\,U^\dag , \A_3) &\;=\;& S(\rho_{\A_3}) \,, \\
    \E(U\,\mathds{1} \otimes \A_2\,U^\dag , \A_3) &\;=\;& 0 \,,
\end{eqnarray}
where $S(\rho) = - \,\rm{Tr}\,\rho\ln\rho$ denotes the von Neumann entropy of the reduced density matrix $\rho \rightarrow \rho_{\A_3}\,$.
\label{lemma:entanglement for GHZ-like states}
\end{lemma}

This means that the maximal possible entanglement with $\A_3$ can be obtained by a subalgebra in $\A_1 \otimes \A_2$ of the same dimension as $\A_3$ whereas the rest is not entangled at all.\\

\noindent\textbf{Proof:}\ \ \
We can write
\begin{equation}\label{omega vector in Schmidt decomposion}
|\,\Omega\,\rangle_{123} \;=\; \sum\limits_{i=1}^d \,\sqrt{\rho_i}\,|\,\varphi_{i} \,\rangle_{12} \otimes |\,\psi_{i} \,\rangle_3 \;,
\end{equation}
where the vector subindices refer to the subalgebras, and $|\,\psi_{i} \,\rangle_3$ represents an ONB for $\A_3$ and $|\,\varphi_{i} \,\rangle_{12}$ is belonging to a basis for $\A_1 \otimes \A_2\,$.

With $|\,\psi_{i} \,\rangle_1$ a basis for $\A_1$ and $|\,\psi_{\alpha} \,\rangle_2$ an arbitrary vector for $\A_2$ we define with help of a unitary transformation
\begin{equation}\label{unitary transform of vector in A1 times A2}
U_{\alpha} |\,\psi_{i} \,\rangle_1 \otimes |\,\psi_{\alpha} \,\rangle_2 \;=\; |\,\varphi_{i} \,\rangle_{12}
\end{equation}
and extend it to a unitary in $\A_1 \otimes \A_2\,$. This serves the purpose. \hspace{0.3cm}\mbox{q.e.d.}\\

It means, as shown before, we can find a unitary transformation such that the states in $\A_3$ are entangled with the states in $\A_1$ in a maximal possible way and separable (product states) with the states in $\A_2\,$. Lemma~\ref{lemma:entanglement for GHZ-like states} is nicely illustrated by Fig.~\ref{fig:concurrence of transformed GHZ theta state} where we can generate entanglement of the traced GHZ state by a unitary transformation up to a maximal concurrence of $C = \frac{1}{2}\,$.

\subsection{Werner states}\label{sec:Werner states}

Next we want to study the Werner states \cite{werner89} as a typical example of mixed states
\begin{equation}
\rho_{\,\rm{Werner}} \;=\; \alpha\, \rho^- \,+\, \frac{1-\alpha}{4}\, \mathds{1}_4 \;=\; \frac{1}{4}
\begin{pmatrix}
1-\alpha & 0        & 0        & 0 \\
0        & 1+\alpha & -2\alpha & 0 \\
0        & -2\alpha & 1+\alpha & 0 \\
0        & 0        & 0        & 1-\alpha \\
\end{pmatrix}\,,
\label{Werner-state in matrix notation}
\end{equation}
or in terms of the Bloch decomposition we have
\begin{equation}\label{Werner-state in Bloch notation}
\rho_{\,\rm{Werner}} \;=\;\frac{1}{4}\left(\mathds{1}\,\otimes\,\mathds{1} \,-\, \alpha\,\vec\sigma\,\otimes\,\vec\sigma\right) \,,
\end{equation}
with the parameter values $\alpha \in [0,1]\,$. They have the interesting feature that they are separable within a certain bound of mixedness, $\alpha \leq 1/3\,$, and within a much larger bound, $\alpha \leq 1/\sqrt{2}\,$, they satisfy a Bell inequality (of CHSH-type) although they contain some amount of entanglement, recall Fig.\ref{fig:tetrahedron of physical states}  for an illustration. Thus the interval $\,\frac{1}{3} < \alpha \leq \frac{1}{\sqrt{2}}\,$ defines the region of local states that are not separable.

Transforming the state $\rho_{\,\rm{Werner}}$ according to Eq.~(\ref{eq:unitary-transform in matrix notation}) we obtain
\begin{eqnarray}
U\,\rho_{\,\rm{Werner}}\,U^\dag &\;=\;& \frac{1}{4}\left(\mathds{1}\,\otimes\,\mathds{1} \,+\, \alpha\,(\sigma_z\,\otimes\,\mathds{1} \,-\, \mathds{1}\,\otimes\,\sigma_z ) \,-\, \sigma_z\,\otimes\,\sigma_z\right) \label{U-transformed Werner-state Bloch notation} \\
&\;=\;& \frac{1}{4}
\begin{pmatrix}
1-\alpha & 0         & 0        & 0 \\
0        & 1+3\alpha & 0        & 0 \\
0        & 0         & 1-\alpha & 0 \\
0        & 0         & 0        & 1-\alpha \\
\end{pmatrix}\,,
\label{U-transformed Werner-state matrix notation}
\end{eqnarray}
which is separable with respect to the algebra $\{ \sigma_i\,\otimes\,\sigma_j \}\,$ for all values of $\alpha \in [0,1]\,$ since the EWI (\ref{def-entwit}) gives (recall the entanglement witness $A_{\rm{opt}}^{\,\rho_{\,\rm{Werner}}} \equiv A_{\rm{opt}}^{\rho^-}$)
\begin{equation}\label{EWI of U-transformed Werner-state}
\left\langle U \rho_{\,\rm{Werner}}\,U^\dag|A_{\rm{opt}}^{\rho^-} \right\rangle \;=\; \textnormal{Tr}\; U \rho_{\,\rm{Werner}}\,U^\dag \,A_{\rm{opt}}^{\rho^-} \;=\; \frac{1}{2\sqrt{3}}(1 - \alpha) \;\geq\; 0 \,.
\end{equation}
This we clearly could expect since the $U$ transformation of the maximal entangled part $\rho_{\,\rm{Werner}}(\alpha =1)=\rho^-$ is already separable.\\

However, transforming also the entanglement witness $U\,A_{\rm{opt}}^{\rho^-}\,U^\dag\,$, Eq.~(\ref{U-transform of EW-Bell minus}), i.e. choosing a different factorization, we then get
\begin{eqnarray} \label{U-transform of EWI Werner-state}
    \left\langle U \rho_{\,\rm{Werner}}\,U^\dag|U A_{\rm{opt}}^{\rho^-}\,U^\dag \right\rangle &\;=\;&
    \left\langle \rho_{\,\rm{Werner}}|A_{\rm{opt}}^{\rho^-} \right\rangle \;=\; \frac{1}{2\sqrt{3}}\,(1 - 3\alpha) \;<\; 0 \,,
\end{eqnarray}
for $\alpha > 1/3\,$, i.e. the transformed Werner state is entangled again with respect to the other algebra factorization $\{ \sigma_i\,\otimes\,\mathds{1}, \mathds{1}\,\otimes\,\sigma_j , \sigma_i\,\otimes\,\sigma_j \}\,$. But as claimed in Theorem~\ref{theorem:factorization-algebra mixed states} the entanglement occurs only beyond a certain bound of mixedness, here for the Werner state the bound is $\alpha > 1/3\,$ .

\subsection{Gisin states}\label{sec:Gisin states}

Whereas Werner \cite{werner89} demonstrated that a mixed entangled state may satisfy a Bell inequality -- thus showing that a Bell inequality is not a complete measure for entanglement -- it was Gisin \cite{gisin} who showed that some quantum states initially satisfying a Bell inequality lead to a violation after certain local selective measurements, i.e. local filtering operations. In this way the nonlocal character of the quantum system is revealed (see also Ref.~\cite{popescu} in this connection). Of course, we can also consider in this case entanglement and separability with respect to the factorization algebra of the density matrix.

Let us begin by introducing the Gisin states \cite{gisin}, they are a mixture of the entangled state $\rho_{\,\theta}$ (\ref{eq:rho theta-state in matrix notation}), discussed before in Chapt.~\ref{sec:Alice-Bob}, and the separable states $\rho_{\uparrow\uparrow}$ and $\rho_{\downarrow\downarrow}$
\begin{equation}\label{Gisin states}
\rho_{\rm{Gisin}}(\lambda ,\theta) \;=\; \lambda \,\rho_{\,\theta} \,+\, \frac{1}{2}(1 - \lambda)(\rho_{\uparrow\uparrow} + \rho_{\downarrow\downarrow}) \,,\qquad \mbox{with} \;\;\;0 \leq \lambda \leq 1\,,
\end{equation}
and in Bloch form they can be written as
\begin{eqnarray}\label{Gisin states in Bloch form}
\rho_{\rm{Gisin}}(\lambda ,\theta) \;=\; & &\frac{1}{4} \big( \,\mathds{1}\,\otimes\,\mathds{1} \,-\,
\lambda \cos (2\theta)\,(\sigma_z\,\otimes\,\mathds{1} \,-\, \mathds{1}\,\otimes\,\sigma_z\,) \nonumber\\
& &\,-\, \lambda \sin (2\theta)\,(\sigma_x\,\otimes\,\sigma_x \,+\, \sigma_y\,\otimes\,\sigma_y )
\,+\, (1 - 2\lambda)\,\sigma_z\,\otimes\,\sigma_z \,\big) \,.
\end{eqnarray}

\noindent Due to a theorem of the Horodeckis \cite{horodecki95} about the maximal violation of a Bell inequality (\`{a} la CHSH \cite{bell-book, chsh69}) we know:
\begin{theorem}[Maximal violation of a Bell inequality \cite{horodecki95}]\ \
Given a general $2 \times 2$ dimensional density matrix in Bloch form $\rho \,=\, \frac{1}{4}\left(\,\mathds{1}\,\otimes\,\mathds{1} + r_i \,\sigma_i\,\otimes\,\mathds{1} +
u_i \,\mathds{1}\,\otimes\,\sigma_i + t_{ij} \;\sigma_i\,\otimes\,\sigma_j\right) \,$ and the Bell operator
\begin{equation}\label{Bell operator}
\B_{\rm{CHSH}} \;=\; \frac{1}{2}\,\big( \vec{a}\cdot\vec{\sigma} \otimes (\vec{b} + \vec{b'})\cdot\vec{\sigma} \;+\; \vec{a'}\cdot\vec{\sigma} \otimes (\vec{b} - \vec{b'})\cdot\vec{\sigma}\big) \,,
\end{equation}
then the maximal violation of the Bell inequality $B^{\rm{max}} \,=\, \rm{max}_{\B}\,\rm{Tr}\,\rho\,\B_{\rm{CHSH}}\,$ is given by
\begin{equation}\label{Bell operator maximal value}
B^{\rm{max}} \;=\; \sqrt{t^2_1 + t^2_2} \;>\; 1 \,,
\end{equation}
where $t^2_1, t^2_2\,$ denote the two larger eigenvalues of the matrices product $(t_{ij})^T (t_{ij})\,$.\\
\label{theorem:horodecki maximal violation of the Bell inequality}
\end{theorem}

\noindent Thus there is a violation if and only if the following parameter condition holds
\begin{equation}\label{Horodecki theorem BI violation}
T(\rho(\lambda ,\theta)) \;=\; \rm{max}\,\{ (2\lambda -1)^2 \,+\, \lambda^2 \sin^2(2\theta), 2\lambda^2 \sin^2(2\theta)\} \;>\; 1 \,.
\end{equation}
Therefore we do not get any violation of the Bell inequality $B \,=\, \rm{Tr}\,\rho_{\rm{Gisin}}\,\B_{\rm{CHSH}} \,\leq\,1\,$ for the parameter range
\begin{equation}\label{lambda bound for BI}
\lambda \;\leq\; \frac{1}{\sqrt{2} \sin(2\theta)}\,,
\end{equation}
assuming $\lambda \leq \frac{1}{2 - \sin (2\theta)}$. \\

\noindent\textbf{Gisin's filtering procedure:}\ \ \
Next Gisin proposes an other type of measurement for the quantum states, a local filtering operation described by the following matrices
\begin{eqnarray}\label{Gisin filter}
F_{\rm{left}} &\;=\;& T_{\rm{left}} \otimes \mathds{1}  \;\quad \mbox{with} \quad\; T_{\rm{left}} \;=\;
\begin{pmatrix}
 \sqrt{\cot \theta} & 0 \\
 0                  & 1 \\
\end{pmatrix}\nonumber\\
F_{\rm{right}} &\;=\;& \mathds{1} \otimes T_{\rm{right}} \quad \mbox{with} \quad T_{\rm{right}} \;=\;\begin{pmatrix}
 1 & 0 \\
 0 & \sqrt{\cot \theta} \\
\end{pmatrix}\,,
\end{eqnarray}
which means that on the left hand side the spin-up component is damped and on the right hand side the spin-down component such that the passing state of the pair  becomes the maximally entangled Bell state $\rho^-\,$. Thus the filtering corresponds to the following map
\begin{eqnarray}\label{filtered Gisin state}
\rho \;\;& \longrightarrow &\;\;
\frac{F_{\rm{left}}\,\rho\,F_{\rm{right}}}{{\rm{Tr}}\,F_{\rm{left}}\,\rho\,F_{\rm{right}}} \;=\; \rho_{\rm{Gisin}}^{\rm{filtered}}\,,\nonumber\\
\rho_{\rm{Gisin}}^{\rm{filtered}}(\lambda ,\theta) &\;=\;& \frac{1}{N}\,\big(\,\lambda \,\sin (2\theta)\,\rho^{-} \,+\, \frac{1}{2}(1 - \lambda)(\rho_{\uparrow\uparrow} + \rho_{\downarrow\downarrow})\,\big) \,,
\end{eqnarray}
with normalization $N = \lambda\,\sin (2\theta) \,+\, (1 - \lambda)\,$. The separable states $\rho_{\uparrow\uparrow} + \rho_{\downarrow\downarrow}\,$ pass the filtering with probability $\cot \theta$ without being absorbed. In Bloch form we have
\begin{eqnarray} \label{filted Gisin state in Bloch decomposition}
\rho_{\rm{Gisin}}^{\rm{filtered}}(\lambda ,\theta) \;=\;
& &\frac{1}{4} \big( \,\mathds{1}\,\otimes\,\mathds{1} \,-\, \frac{\lambda\,\sin (2\theta)}{1 + \lambda\,\sin (2\theta)}\,(\sigma_x\,\otimes\,\sigma_x \,+\, \sigma_y\,\otimes\,\sigma_y)\nonumber\\
& &\,-\, \frac{1 - \lambda - \lambda\,\sin (2\theta)}{1 - \lambda + \lambda\,\sin (2\theta)}\,\sigma_z\,\otimes\,\sigma_z \,\big) \,.
\end{eqnarray}

Then Horodecki's Theorem~\ref{theorem:horodecki maximal violation of the Bell inequality} implies that there is a violation of the Bell inequality for the parameter values
\begin{equation}\label{lambda bound for BI for filtered Gisin state}
\lambda \;>\; \frac{1}{1 + \sin(2\theta)(\sqrt{2} - 1)}\,.
\end{equation}

\begin{figure}
	\centering
	\includegraphics[width=0.55\textwidth]{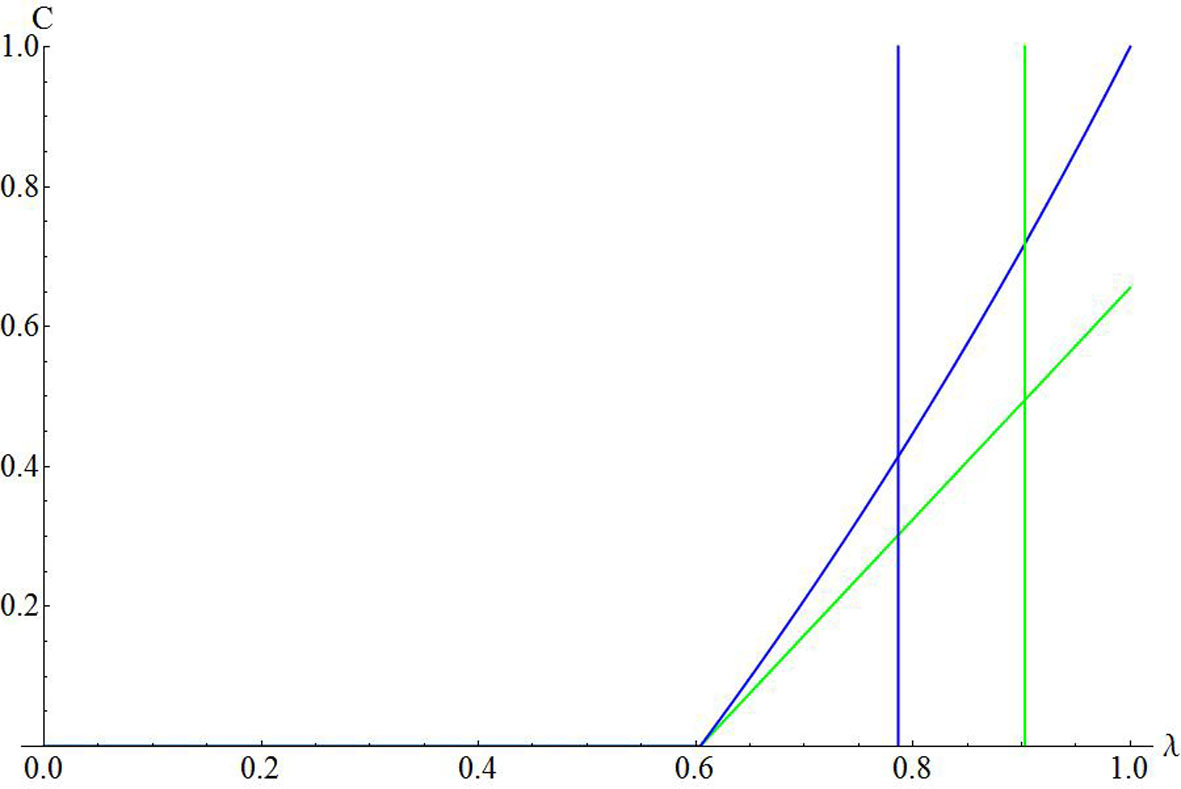}
	\caption{The concurrence of the Gisin state with $\theta = 0.35$ (green line) is plotted in dependence of the parameter $\lambda$ and the concurrence of the filtered Gisin state (lila line) together with their Bell inequality bounds, the corresponding vertical lines. Violation of the Bell inequality occurs on the right hand side of the vertical line.}
	\label{fig:concurrence_Gisin_BI-violation}
\end{figure}

The local filtering increases the amount of entanglement such that the Bell inequality is violated, see Fig.~\ref{fig:concurrence_Gisin_BI-violation}.
We have plotted the concurrence of the Gisin state with $\theta = 0.35$ (green line) in dependence of the parameter $\lambda$ and the concurrence of the filtered Gisin state (lila line) together with their Bell inequality bounds, the corresponding vertical lines. The violation of the Bell inequality occurs on the right hand side of the vertical line. We see that for $\lambda \leq 0.9$ the Gisin state satisfies the Bell inequality (green lines) whereas the filtered state violates the inequality already for $\lambda > 0.78\,$ (lila lines).

Thus for Gisin states there are values of the parameters $\lambda$ and $\theta$ such that the state $\rho_{\rm{Gisin}}(\lambda ,\theta)$ is local (in the sense of satisfying a Bell inequality) but the corresponding filtered state $\rho_{\rm{Gisin}}^{\rm{filtered}}(\lambda ,\theta)$ is not (i.e. violates a Bell inequality).\\

The above described procedure is certainly different to our view of the free choice of factorizing the algebra of a density matrix. Gisin uses \emph{nonunitary} but local filtering operations which increase the nonlocal quantum correlations of a system. In contrast we work with \emph{unitary} but nonlocal operations to switch between the different factorizations of the Hilbert-Schmidt space where a given state appears either separable or entangled, depending on our free choice. The mixedness of the quantum states changes in Gisin's filtering procedure, in our operations not.

\begin{figure}
	\centering
	\includegraphics[width=0.55\textwidth]{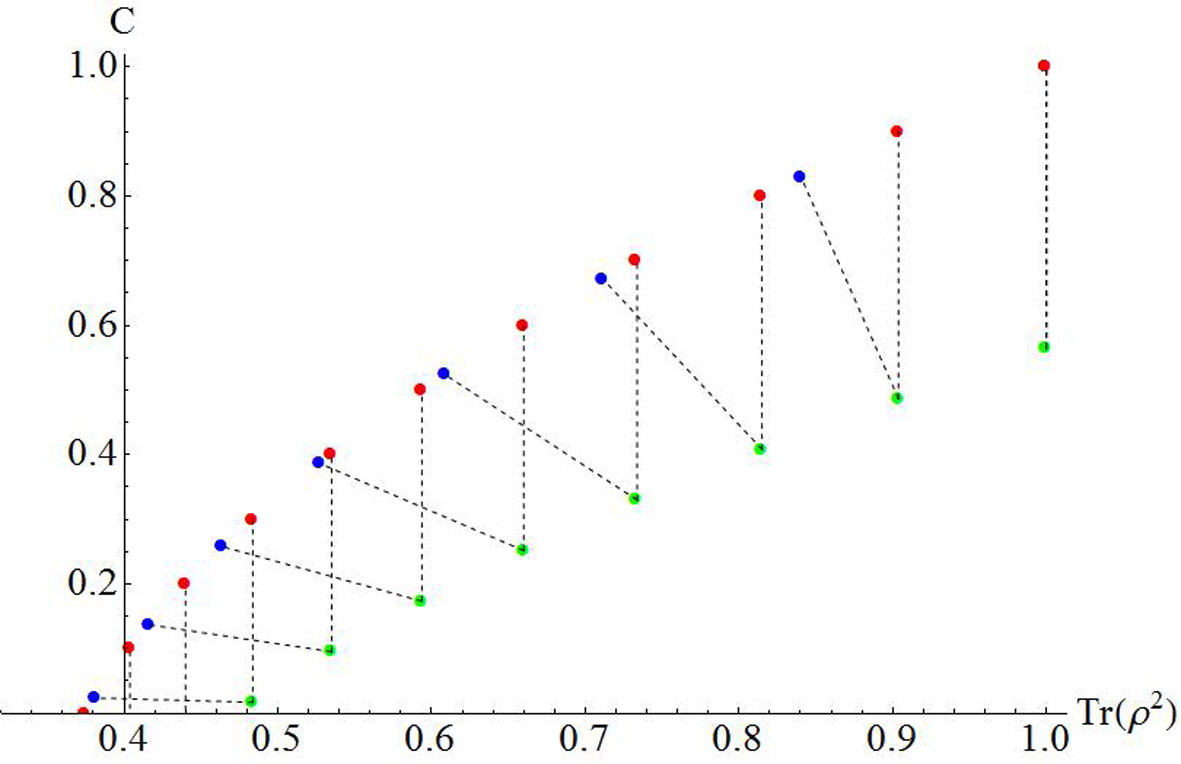}
	\caption{The concurrence is plotted versus the purity $P(\rho) = \textnormal{Tr}\,\rho^2$ of the quantum states, for the Gisin states (green dots), the filtered Gisin states (blue dots) and the unitary transformed Gisin states (red dots). The lines connect the states of the same value of the parameter $\lambda$.}
	\label{fig:concurrence-mixedness Gisin states}
\end{figure}

When we transform the Gisin state $\rho_{\rm{Gisin}}(\lambda ,\theta)$ with our unitary transformation $U_\theta$ (\ref{eq:unitary-theta-transform in matrix notation})
\begin{equation}\label{unitarily transformed Gisin states}
\rho_{\rm{Gisin}}^{\rm{unitary}}(\lambda) \;=\; U_\theta\,\rho_{\rm{Gisin}}(\lambda ,\theta)\,U_\theta^\dagger \;=\; \lambda \,\rho^{+} \,+\, \frac{1}{2}(1 - \lambda)(\rho_{\uparrow\uparrow} + \rho_{\downarrow\downarrow}) \,,
\end{equation}
we achieve a constant amount of entanglement for all $\theta$ values depending only on the parameter $\lambda\,$. We have compared the two procedures by calculation the concurrence versus the purity $P(\rho) = \textnormal{Tr}\,\rho^2$. The results we have plotted on Fig.\ref{fig:concurrence-mixedness Gisin states}.
Whereas Gisin's filtering procedure (blue dots) increases the concurrence and decreases the purity of the origin Gisin states (green dots) such that the Bell inequality is violated, our unitary operations (red dots) just increase the concurrence keeping the purity fixed leading to a higher value of the violation of the Bell inequality. The lines connect the states of the same value of the parameter $\lambda$. For example, for $\theta = 0.35$ and $\lambda = 0.8$ the values of the concurrence are: $C = 0.4$ for $\rho_{\rm{Gisin}}$, $C = 0.62$ for the filtered state $\rho_{\rm{Gisin}}^{\rm{filtered}}$ and $C = 0.75$ for the unitarily transformed state $\rho_{\rm{Gisin}}^{\rm{unitary}}$.

\begin{figure}
	\centering
	\includegraphics[width=0.55\textwidth]{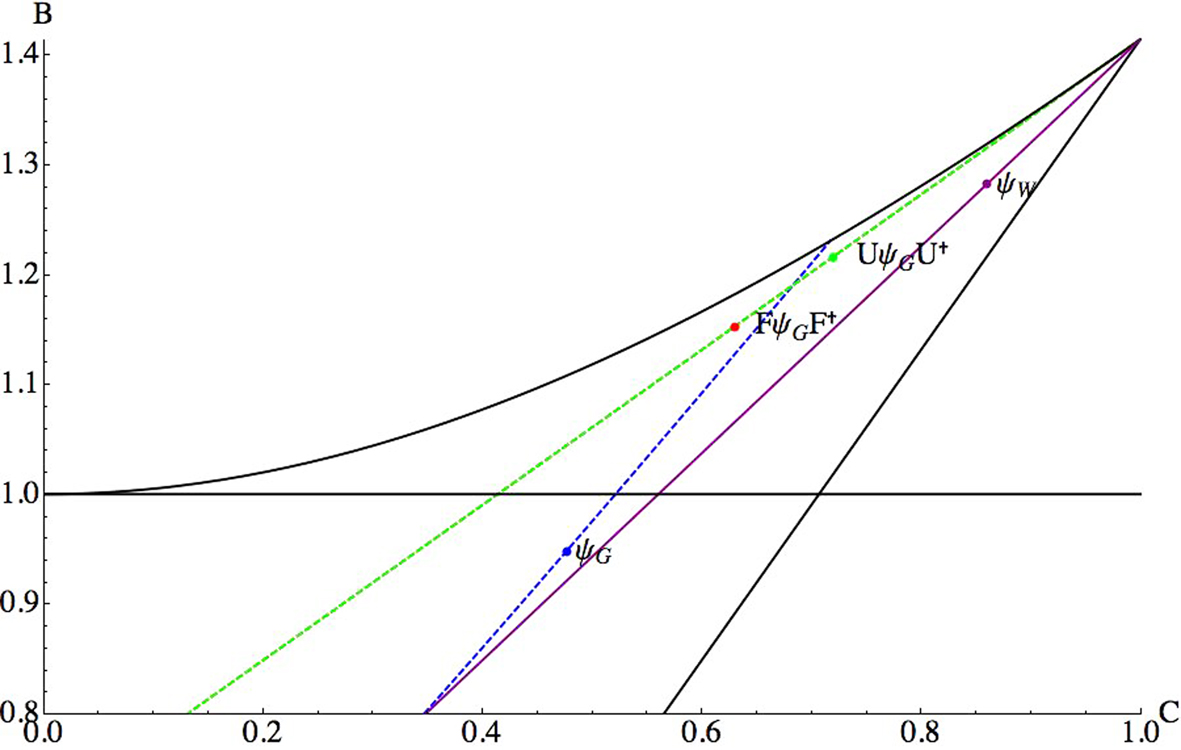}
	\caption{The Bell violation $B$ together with the Bell bound $B = 1$ is plotted versus the concurrence $C$ for the several types of quantum states. The black curves constitute the Verstraete-Wolf bounds, on the green line the filtered and unitarily transformed Gisin states are located. The blue and lila lines represent the Gisin and Werner states.}
	\label{fig:Bell violation and Verstraete-Wolf bounds}
\end{figure}

For a given concurrence the amount of the violation of the Bell inequality is bounded from above $B^{\rm{upper}} = \sqrt{1 + C^2}$ and below $B^{\rm{lower}} = \rm{max}(1,\sqrt{2}C)$ for all quantum states due to the Verstraete-Wolf Theorems \cite{Verstraete-Wolf}, see Fig.~\ref{fig:Bell violation and Verstraete-Wolf bounds}.
The black curves constitute the Verstraete-Wolf bounds, on the green line the filtered and unitarily transformed Gisin states are located. For example, for $\lambda = 0.86$ and $\theta = 0.4$ the filtered state corresponds to the red dot and the unitarily transformed one to the green dot. The blue and lila lines represent the Gisin and Werner states, the Gisin state for above parameter values is well below the Bell bound $B = 1\,$ located, the Werner state lies above.

However, if we also transform the Bell operator then, of course, the Bell bound remains invariant for the transformed and untransformed case
\begin{equation}\label{unitary invariance of Bell inequality}
B \;=\; {\rm{Tr}}\,\rho_{\rm{Gisin}}^{\rm{unitary}}\,U_\theta\,\B_{\rm{CHSH}}\,U_\theta^\dagger
\;=\; \rm{Tr}\,\rho_{\rm{Gisin}}\,\B_{\rm{CHSH}} \,.\\
\end{equation}

\vspace{0.3cm}

\begin{figure}
	\centering
	\includegraphics[width=0.55\textwidth]{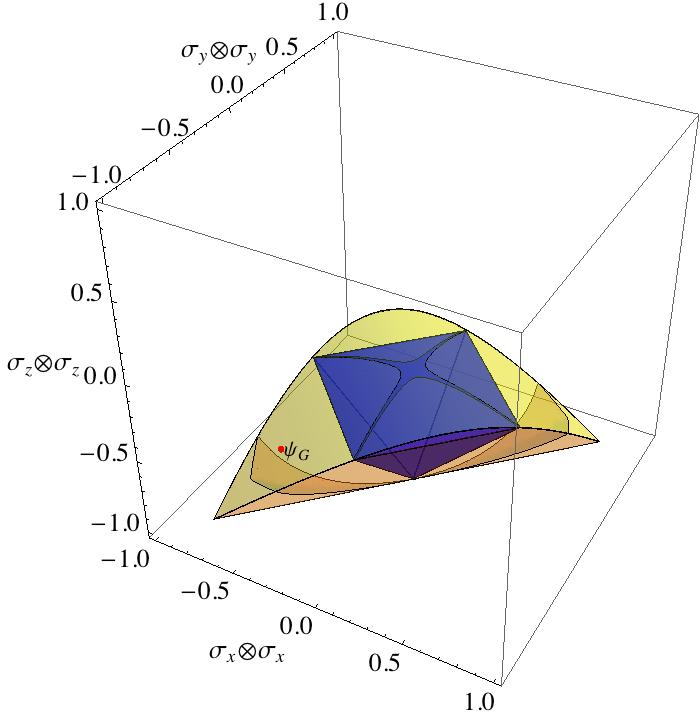}
	\caption{Shrunk tetrahedron with Gisin state. Due to the occurrence of the term ($\sigma_z\,\otimes\,\mathds{1} \,-\, \mathds{1}\,\otimes\,\sigma_z$) in the density matrix the former tetrahedron of Weyl states shrinks from above becoming parabolic in z-direction. The Gisin state $\psi_{\rm{G}}$ (\ref{Gisin states}) with parameter values $\theta = 0.35$ and $\lambda = 0.8$ is located well within the Bell bound (dark-yellow surface).}
	\label{fig:shrinked tetrahedron with Gisin state}
\end{figure}

\noindent\textbf{Geometry of the quantum states:}\ \ \
It's illustrative to demonstrate the geometry of the above described quantum states. The occurrence of the term ($\sigma_z\,\otimes\,\mathds{1} \,-\, \mathds{1}\,\otimes\,\sigma_z$) in the density matrix of the Gisin states (\ref{Gisin states}) implies a shrinkage from above of the former tetrahedron of Weyl states, which becomes parabolic in z-direction, see Fig.~\ref{fig:shrinked tetrahedron with Gisin state}. The Gisin states constitute a curve which lies on the surface of the shrunk tetrahedron connecting the bottom with the top. The bound of the Bell operator is given by the dark-yellow surfaces and the separable states constitute the shrunk double pyramid (shaded in blue). We see that the Gisin state (\ref{Gisin states}), for example, for $\theta = 0.35$ and $\lambda = 0.8$ is, although entangled, well within the Bell bound, the region of local states.

\begin{figure}
	\centering
	\includegraphics[width=0.55\textwidth]{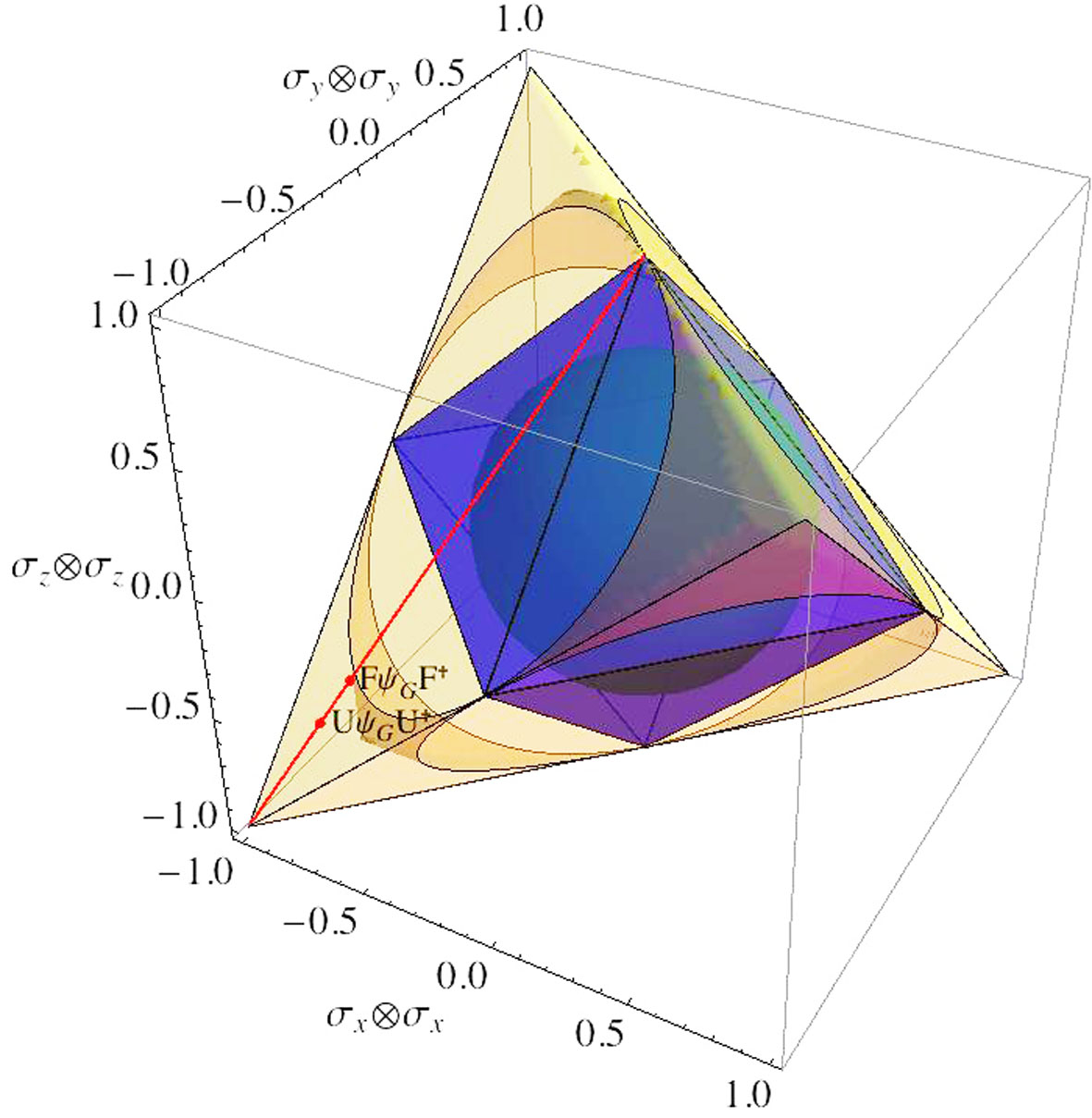}
	\caption{Tetrahedron of physical states. The Gisin line (in red) reaches from the maximal entangled Bell state $\rho^-$ on the bottom to the top separable state of double pyramid, represented by the mixture $\rho_{\uparrow\uparrow} + \rho_{\downarrow\downarrow}$. The filtered $F\psi_{\rm{G}}F^{\dagger}$ and the unitarily transformed $U\psi_{\rm{G}}U^{\dagger}$ Gisin states are plotted for $\lambda = 0.8\,$ and lie outside the Bell bound (dark-yellow surface).}
	\label{fig:tetrahedron with filtered and unitarily transformed states}
\end{figure}

On the other hand, all unitarily transformed (\ref{unitarily transformed Gisin states}) and filtered (\ref{filtered Gisin state}) Gisin states lie on a line -- the Gisin line (in red) -- between the maximal entangled Bell state $\rho^-$ on the bottom and the separable mixture $\rho_{\uparrow\uparrow} + \rho_{\downarrow\downarrow}$ on top of the double pyramid, see Fig.~\ref{fig:tetrahedron with filtered and unitarily transformed states}.
We see that both kind of states for the parameter value $\lambda = 0.8\                                                                           ,$ lie outside the Bell bound, i.e. they violate the Bell inequality. The filtered state (\ref{filtered Gisin state}) violates the Bell inequality less, is nearer to the tracial state at origin, since the filtering increases its mixedness, whereas the unitarily transformed state (\ref{unitarily transformed Gisin states}) keeps the mixedness constant and therefore violates the Bell inequality more.

The Gisin line composed of the maximal entangled state $\rho^-$ and the orthogonal separable state $\rho_{\uparrow\uparrow} + \rho_{\downarrow\downarrow}$, see Fig.~\ref{fig:tetrahedron with filtered and unitarily transformed states}, is a nice example for Lemma~\ref{lemma:bound for werner state}, where the bound $\beta > \frac{1}{2}$ for entanglement is indeed optimal and corresponds to the red line part in the yellow region.

\section{Factorization in physical examples}\label{sec:factorization in physical examples}

\subsection{Quantum teleportation}\label{sec:quantum teleportation}

Quantum teleportation \cite{BBCJPW} and its experimental verification \cite{bouwmeester-pan-zeilinger-etal, ursin-zeilinger-etal} became in the recent years a popular subject in quantum information. It is an amazing quantum feature which lives from the fact that several qubits can be entangled in different ways. Usually three qubits are considered together with the associated Bell states. We don't want to repeat here the usual treatment but wish to explore the essential features of maximally entangled states, which lead to quantum teleportation.

We study the tensor product of three matrix algebras $\A_1 \otimes \A_2 \otimes \A_3$ of equal dimensions, $\A_1, \A_2$ belonging to Alice and $\A_3$ to Bob. The situation is that Alice gets on her first line $\A_1$ an incoming message given by a vector $|\phi \rangle$ which she wants to transfer to Bob without direct contact between the algebras $\A_1$ and $\A_3$, though she knows what the corresponding vectors are. To achieve this goal she uses the fact that the three algebras can be entangled in different ways. Alice also knows that her second line $\A_2$ is entangled with Bob via an EPR source, such that the total state restricted to $\A_2 \otimes \A_3$ is maximally entangled. A maximally entangled state $|\psi \rangle_{23} \in \A_2 \otimes \A_3$ defines an isometry $I_{23}$ (a bijective map that preserves the distances) between the vectors of one factor to the other.

The possibility to transfer the incoming state of $\A_1$ at Alice into a state of $\A_3$ at Bob uses the fact that an isometry $I_{13}$ between this two algebras was taken for granted. Now Alice chooses an isometry $I_{12}\,$, which correspond to choosing a maximally entangled state $|\psi \rangle_{12} \in \A_1 \otimes \A_2\,$, such that the following isometry relation holds
\begin{equation}\label{isometry relation for 1-3}
I_{12} \cdot I_{23} \;=\; I_{13}\,.
\end{equation}
Expressed in an ONB $\{ \varphi_i \}$ of one factor the state vector can be written as
\begin{equation}\label{psi-12 isometry}
|\,\psi\,\rangle_{12} \;=\; \frac{1}{\sqrt{d}}\,\sum\limits_{i=1}^d \,|\,\varphi_{i} \,\rangle_1 \otimes | \,I_{12} \,\varphi_{i} \,\rangle_2 \;.
\end{equation}
A measurement by Alice in $\A_1 \otimes \A_2$ with the outcome of this entangled state produces the desired state vector in $\A_3\,$. The outcome of other maximally entangled states, orthogonal to the first one, corresponds to a unitary transformation $U_{12}$ in $\A_1 \otimes \A_2\,$, which produces a unique unitary transformation $U_{3}\,$ in $\A_3$ that Bob can perform to obtain the desired state. Thus Alice just has to tell Bob her measurement outcome via some classical channel. The measurements of Alice produce the following results for Bob:
\begin{eqnarray}\label{eq:measurements of Alice in teleportation}
\big(\left|\,\psi\,\right\rangle\left\langle\,\psi\,\right|\big)_{12} \otimes \mathds{1}_{3} \;|\,\phi\,\rangle_1 \otimes |\,\psi\,\rangle_{23} &\;=\;& \frac{1}{d^2}\,|\,\psi\,\rangle_{12} \otimes |\,\phi\,\rangle_{3} \,,\\
U_{12}\big(\left|\,\psi\,\right\rangle\left\langle\,\psi\,\right|\big)_{12}\,U_{12}^{\dagger}\, \otimes \mathds{1}_{3} \;|\,\phi\,\rangle_1 \otimes |\,\psi\,\rangle_{23} &\;=\;& \frac{1}{d^2}\,U_{12}|\,\psi\,\rangle_{12} \otimes U_{3}|\,\phi\,\rangle_{3} \;,
\end{eqnarray}
where the state vectors can be expressed in an ONB for a fixed isometry, e.g. for $I_{12} = \mathds{1}$, by
\begin{eqnarray}\label{state vectors in teleportation}
|\,\psi\,\rangle_{23} &\;=\;& \frac{1}{\sqrt{d}}\,\sum\limits_{i=1}^d\,|\,\varphi_{i} \,\rangle_2 \otimes |\,\varphi_{i} \,\rangle_3 \;,\\
|\,\phi\,\rangle_{1 \,\rm{or}\, 3} &\;=\;& \sum\limits_{i=1}^d\,\alpha_{i}\,|\,\varphi_{i} \,\rangle_{1 \,\rm{or}\, 3} \;.
\end{eqnarray}

Summarizing, if Alice measures the same entanglement between $\A_1$ and $\A_2$ as there was between $\A_2$ and $\A_3$, which was given by the EPR source, she knows that her measurement left Bob's $\A_3$ in Alice's incoming state $|\phi \rangle_3\,$. If Alice finds, on the other hand, a different Bell state, which is given by $U_{12}|\psi \rangle_{12}$ since all other Bell states are connected by unitary transformations, then Bob will have the state vector $U_{3}|\phi \rangle_{3}$, where the unitary transformation $U_{3}$ is determined by $U_{12}\,$.

Note, our view of interpreting the maximal entangled states as isometries between the vectors of one algebra to the other has the merit of being quite general. It is independent of the special choice of coordinates or vectors and it works in \emph{any} dimension. Loosely speaking, quantum teleportation relies on the fact that we may cut a cake in different ways (factors).

\subsection{Entanglement swapping}\label{sec:entanglement swapping}

Closely related to the teleportation of single quantum states is an other striking quantum phenomenon called entanglement swapping \cite{zukowski-zeilinger-horne-ekert}, it is the teleportation of entanglement. Experimentally entanglement swapping has been demonstrated in Ref.~\cite{pan-bouwmeester-weinfurter-zeilinger} and is nowadays a standard tool in quantum information processing \cite{bertlmann-zeilinger02}. It illustrates the different slicing of a 4-fold tensor product into $(1,2) \otimes (3,4)$ or $(1,4) \otimes (2,3)\,$, where e.g. $(1,2)$ denotes entanglement between the subsystems $\A_1$ and $\A_2\,$.

The same way of reasoning as in Sec.~\ref{sec:quantum teleportation} can be applied to entanglement swapping, i.e. we consider a maximally entangled state as an isometry between the vectors of one factor to the other. In case of entanglement swapping the starting point are two maximally entangled pure states combined in a tensor product $|\,\psi\,\rangle_{12} \otimes |\,\psi\,\rangle_{34}\,$. Expressed in an ONB of one factor the entangled states are given by Eq.~(\ref{psi-12 isometry}). They describe usually two pairs of EPR photons. These propagate into different directions and at the interaction point of two of them, say photon $2$ and $3$, a so-called Bell state measurement is performed, i.e. a measurement with respect to an orthogonal set of maximally entangled states. In the familiar case they are given by the Bell states $\left|\,\psi^{\,\pm}\,\right\rangle_{23} , \left|\,\phi^{\,\pm}\,\right\rangle_{23}\,$. But our results are more general, the four factors just have to be of the same dimension $d$ which is arbitrary.

In complete analogy to the case of teleportation, discussed before, the effect of the projection corresponding to the measurement on the state is
\begin{equation}\label{eq:measurement for 2-3 in entanglement swapping}
\big(\left|\,\psi\,\right\rangle\left\langle\,\psi\,\right|\big)_{23} \otimes \mathds{1}_{14} \;|\,\psi\,\rangle_{12} \otimes |\,\psi\,\rangle_{34} \;=\; \frac{1}{d^2}\,|\,\psi\,\rangle_{23} \otimes |\,\psi\,\rangle_{14} \;,
\end{equation}
where $|\psi \rangle_{14}$ is a maximally entangled state corresponding to the isometry $I_{14}$ which satisfies the relation
\begin{equation}\label{isometry relation for 1-4}
I_{14} \;=\; I_{12} \cdot I_{23} \cdot I_{34} \;.
\end{equation}
The other isometries $I_{12}, I_{23}, I_{34}\,$ correspond to the other maximally entangled states.

Thus after the Bell state measurement of photon $2$ and $3$ into a definite entangled state the photons $1$ and $4$ become instantaneously entangled into the same state, remarkably, the two photons originate from different noninteracting sources.

This effect of entanglement swapping can also be interpreted as teleportation of an unknown state of photon $2$ onto photon $4\,$. Thus the teleported photon has no well defined polarization. The reason is that a unitary transformation of the Bell state measurement expressed by $U_2 \cdot I_{23}\,$ creates a unitary transformation $U_1 |\psi \rangle_{14}\,$ on the remaining state with the same unitary matrix.

\section{Discussion and Conclusion} \label{sec:conclusion}

First of all, we want to emphasize that the different factorization algebras of a density matrix, corresponding to a quantum state, are not at all unique. They can, however, be chosen in order to illustrate in a natural way the physical interpretation. Clearly, for an experimentalist the factorization normally considered is fixed by the set-up, however, for an absent-minded theorist an arbitrary factorization of the algebra seems to be natural to play with and leads to different results of entanglement. We have investigated how different these results can be. For pure states the situation is quite clear, we can always switch between separability and maximal entanglement. However, for mixed states a minimal mixedness is required because the tracial state and a sufficiently small neighborhood is separable for any factorization.\\

We should point out that the question how to factorize a given algebra appears in many considerations related to concrete physical situations. Let us mention some examples.

Think of the Hanbury Brown Twiss effect \cite{HanburyBrownTwiss}, where photons are produced so far apart that most certainly they are not entangled. Nevertheless, they are able to produce non-local correlations in joint measurement experiments. Taking into account that every mixed state can be considered to be pure on a large algebra non-local correlations correspond to entanglement for appropriate subalgebras that, e.g., reflect Bose or Fermi statistics, bunching or antibunching effects in the corresponding experiments \cite{henny-etal99, bromberg-etal2010}.

In particle physics, as an other example, the neutral K-mesons can be considered as \emph{kaonic qubits} \cite{bertlmann-hiesmayr2006}. They can also be analyzed with respect to entanglement, where the subalgebras are determined by the fact that we concentrate either on the production of the kaons, i.e. on the strangeness states $K^0 \bar{K}^0\,$, or on their decays, i.e. on their short- and long-lived states $K_S K_L\,$.

Especially in relativistic quantum field theory it is important to be precise, what are the chosen subalgebras if one talks about entanglement. Local subalgebras (i.e. double cones) are always entangled due to the Reeh-Schlieder Theorem \cite{reeh-schlieder, haag-book}. It implies that the vacuum state is not positive under partial transposition and cannot be separable \cite{narnhofer-PLA2003}. However, the local algebras are so large that correlations corresponding to the entanglement may be hidden for the observer and therefore cannot be used as source for observable effects. We have to choose smaller algebras corresponding to some modes. But here we have to be careful that these subalgebras can be controlled by the experimentalist. In particular, acceleration of the observer can change the amount of observable entanglement \cite{fuentes-mann, alsing-fuentes-etal, friis-b-h-h, friis-k-m-b, smith-mann}. On the other hand, it can be important to restrict to separable states that are not influenced by the environment. Here we are concerned with the problem, how far it is possible to find localized algebras with vanishing entanglement \cite{narnhofer-ClassQuantGrav2011, narnhofer-tobepub}.

Another example is provided by particles in a constant magnetic field reduced to two dimensions. It relates to the Quantum Hall Effect \cite{laughlin1981}, where it is essential to combine gauge independence and Fermi statistics, that always asks for a kind of entanglement. Here the groundstate and eigenstates of the Hamiltonian are known but gauge dependent. Subalgebras should be constructed in a way that they are gauge invariant so that the notion of entanglement remains physically meaningful \cite{narnhofer-RepMathPhys2004}.

As a last problem, that appears quite naturally, we want to mention a generalization of Theorem~\ref{theorem:factorization-algebra constrained}. Let us consider a state with some uncertainty, that means we do not vary over all unitary transformations of the state but just over a subclass that, e.g., reflects the coupling to an environment. What can we say about the possible purity, and even more, about the possible entanglement under this restriction?\\

In this Article we analyzed our results in detail for qubits, the familiar case of Alice \& Bob in quantum information, and demonstrated explicitly how we can switch between separability and entanglement. We discussed our general statements in particular for the GHZ states, the Werner states and the Gisin states by showing concretely the effect of the unitary switch, which in the later case differs from experimental local filtering operations.

From the many phenomena, where this unitary switch between separability and entanglement is crucial, we just picked out two of them, namely quantum teleportation and entanglement swapping. We pointed out that the experimental result is based on entanglement of different factorizations. Therefore, speaking of entanglement without specifying the factorization of the total algebra corresponding to the quantum  state does not make sense. In our argumentation we concentrated on the fact that entanglement of \emph{pure} states defines a natural isometry between the partners and therefore can easily be extended to several partners without any restrictions to dimensions.

Finally, our goal has been to find the right frame of mind to digest the richness of the familiar physical results.

\begin{acknowledgments}

One of the authors (W. T.) is grateful to Peter Zoller for the inspiring Lecture \cite{zoller_ESI-lecture} which made him realize that the entanglement of a state without further ado does not make sense.

\end{acknowledgments}


\begin{thebibliography}{100}

\bibitem{EPR}
A. Einstein, B. Podolsky, and N. Rosen, Phys. Rev. \textbf{47}, 777 (1935)

\bibitem{schrodinger1935}
E. Schr\"odinger, Naturwissenschaften \textbf{23}, 807-812, 823-828, 844-849 (1935)

\bibitem{bell1964}
J. S. Bell, Physics \textbf{1}, 195 (1964)

\bibitem{clauser-freedman1972}
S. J. Freedman and J. F. Clauser, Phys. Rev. Lett. \textbf{28}, 938 (1972)

\bibitem{clauser1976}
J. F. Clauser, Phys. Rev. Lett. \textbf{36}, 1223 (1976)

\bibitem{aspect-grangier-roger1982}
A. Aspect, P. Grangier, and G. Roger, Phys. Rev. Lett. \textbf{49}, 91 (1982)

\bibitem{aspect-dalibard-roger1982}
A. Aspect, J. Dalibard, and G. Roger, Phys. Rev. Lett. \textbf{49}, 1804 (1982)

\bibitem{weihs-zeilinger1998}
G. Weihs, T. Jennewein, C. Simon, H. Weinfurter, and A. Zeilinger, Phys. Rev. Lett. \textbf{81}, 5039 (1998)

\bibitem{bertlmann-zeilinger02}
R. A. Bertlmann and A. Zeilinger (eds.), \emph{Quantum [Un]speakables}, Springer 2002

\bibitem{bowmeester-zeilinger2000}
D. Bouwmeester, A. Ekert, and A. Zeilinger (eds.), \emph{The physics of quantum
information: quantum cryptography, quantum teleportation, quantum computations},
Springer 2000.

\bibitem{BBCJPW}
C. H. Bennett, G. Brassard, C. Cr\'epeau, R. Jozsa, A. Peres, and W.K. Wootters, Phys. Rev. Lett. \textbf{70}, 1895 (1993)

\bibitem{bouwmeester-pan-zeilinger-etal}
D. Bouwmeester, J. W. Pan, K. Mattle, M. Eibl, H. Weinfurter, and A. Zeilinger, Nature \textbf{390}, 575 (1997)

\bibitem{ursin-zeilinger-etal}
R. Ursin, T. Jennewein, M. Aspelmeyer, R. Kaltenbaek, M. Lindenthal, P. Walther, and A. Zeilinger, Nature \textbf{430}, 849 (2004)

\bibitem{zanardi-lidar-lloyd2004}
P. Zanardi, D. A. Lidar, and S. Lloyd, Phys. Rev. Lett. \textbf{92}, 060402 (2004)

\bibitem{zanardi2001}
P. Zanardi, Phys. Rev. Lett. \textbf{87}, 077901 (2001)

\bibitem{harshman-ranade}
N. L. Harshman and K.S. Ranade, Phys. Rev. A \textbf{84}, 012303 (2011)

\bibitem{lewenstein-Bruss-etal}
M. Lewenstein, D. Bru\ss, J. I. Cirac, B. Kraus, M. Ku\'s, J. Samsonowicz, A. Sanpera, and R. Tarrach, J. Mod. Opt. \textbf{47}, 2841 (2000)

\bibitem{vollbrecht-werner00}
K. G. H. Vollbrecht and R. F. Werner, J. Math. Phys. \textbf{41}, 6772 (2000)

\bibitem{werner01}
R. F. Werner, J. Phys. A: Math. Gen. \textbf{34}, 7081 (2001)

\bibitem{narnhofer06}
H. Narnhofer, J. Phys. A: Math. Gen. \textbf{39}, 7051 (2006)

\bibitem{bertlmann-krammer-AnnPhys09}
R. A. Bertlmann and P. Krammer, Ann. Phys. \textbf{324}, 1388 (2009)

\bibitem{baumgartner-hiesmayr-narnhofer06}
B. Baumgartner, B. C. Hiesmayr, and H. Narnhofer, Phys. Rev. A \textbf{74}, 032327 (2006)

\bibitem{baumgartner-hiesmayr-narnhofer07}
B. Baumgartner, B. C. Hiesmayr, and H. Narnhofer, J. Phys. A: Math. Theor. \textbf{40}, 7919 (2007)

\bibitem{baumgartner-hiesmayr-narnhofer08}
B. Baumgartner, B. C. Hiesmayr, and H. Narnhofer, Phys. Lett. A \textbf{372}, 2190 (2008)

\bibitem{bertlmann-krammer-JPA08}
R. A. Bertlmann and P. Krammer, J. Phys. A: Math. Theor. \textbf{41}, 235303 (2008)

\bibitem{horodecki99}
P. Horodecki, M. Horodecki, and R. Horodecki, Phys. Rev. Lett. \textbf{82}, 1056 (1999)

\bibitem{bertlmann-krammer-PRA-78-08}
R. A. Bertlmann and P. Krammer, Phys. Rev. A \textbf{78}, 014303 (2008)

\bibitem{bertlmann-krammer-PRA-77-08}
R. A. Bertlmann and P. Krammer, Phys. Rev. A \textbf{77}, 024303 (2008)

\bibitem{verstraete-audenaert-demoor}
F. Verstraete, K. Audenaert, and B. DeMoor, Phys. Rev. A \textbf{64}, 012316 (2001)

\bibitem{peres96}
A. Peres, Phys. Rev. Lett. \textbf{77}, 1413 (1996)

\bibitem{horodecki96}
M. Horodecki, P. Horodecki, and R. Horodecki, Phys. Lett. A \textbf{223}, 1 (1996)

\bibitem{kus-zyczkowski}
M. Ku\'s and K. \.Zyczkowski, Phys. Rev. A \textbf{63}, 032307 (2001)

\bibitem{zyczkowski-bengtsson}
K. \.Zyczkowski and I. Bengtsson, arXiv:quant-ph/0606228

\bibitem{bengtsson-zyczkowski-book}
I. Bengtsson and K. \.Zyczkowski, \emph{Geometry of Quantum states: An Introduction to Quantum Entanglement}, Cambridge University Press 2006

\bibitem{gurvits-barnum}
L. Gurvits and H. Barnum, Phys. Rev. A \textbf{66}, 062311 (2002)

\bibitem{ishizaka-hiroshima}
S. Ishizaka and T. Hiroshima, Phys. Rev. A \textbf{62}, 022310 (2000)

\bibitem{bertlmann-narnhofer-thirring02}
R. A. Bertlmann, H. Narnhofer, and W. Thirring, Phys. Rev. A \textbf{66}, 032319 (2002)

\bibitem{vollbrecht-werner-PRA00}
K. G. H. Vollbrecht and R. F. Werner, Phys. Rev. A \textbf{64}, 062307 (2000)

\bibitem{horodecki-R-M96}
R. Horodecki and M. Horodecki, Phys. Rev. A \textbf{54}, 1838 (1996)

\bibitem{spengler-huber-hiesmayr09}
C. Spengler, M. Huber, and B. C. Hiesmayr, J. Phys. A: Math. Theor. \textbf{44}, 065304 (2011)

\bibitem{munro-james-white-kwiat}
W. J. Munro, D. F. V. James, A. G. White, and P.G. Kwiat, Phys. Rev. A \textbf{64}, 030302 (2001)

\bibitem{terhal00}
B. M. Terhal, Phys. Lett. A \textbf{271}, 319 (2000)

\bibitem{bruss02}
D. Bru\ss, J. Math. Phys. \textbf{43}, 4237 (2002)

\bibitem{wootters98}
W. K. Wootters, Phys. Rev. Lett. \textbf{80}, 2245 (1998)

\bibitem{hill-wootters97}
S. Hill and W. K. Wootters, Phys. Rev. Lett. \textbf{78}, 5022 (1997)

\bibitem{wootters01}
W. K. Wootters, Quantum Information and Computation \textbf{1}, 27 (1998)

\bibitem{GHZ89}
D.M. Greenberger, M.A. Horne, and A. Zeilinger, Going beyond Bell's theorem, in \emph{Bell's Theorem, Quantum Theory, and Conceptions of the Universe}, M. Kafatos (ed.), Kluwer Dortrecht 1989, pp 73-76

\bibitem{Pan-etal00}
J.-W. Pan, D. Bouwmeester, M. Daniell, H. Weinfurter, and A. Zeilinger, Nature \textbf{403}, 515 (2000)

\bibitem{werner89}
R. F. Werner, Phys. Rev. A \textbf{40}, 4277 (1989)

\bibitem{gisin}
N. Gisin, Phys. Lett. A \textbf{210}, 151 (1996)

\bibitem{popescu}
S. Popescu, Phys. Rev. Lett. \textbf{74}, 2619 (1995)

\bibitem{horodecki95}
R. Horodecki, P. Horodecki, and M. Horodecki, Phys. Lett. A \textbf{200}, 340 (1995)

\bibitem{bell-book}
J. S. Bell, \emph{Speakable and unspeakable in quantum mechanics}, Cambridge University Press 1987

\bibitem{chsh69}
J. F. Clauser, M. A. Horne, A. Shimony, and R. A. Holt, Phys. Rev. Lett. \textbf{23}, 880 (1969)

\bibitem{Verstraete-Wolf}
F. Verstraete and M.M. Wolf, Phys. Rev. Lett. \textbf{89}, 170401 (2001)

\bibitem{zukowski-zeilinger-horne-ekert}
M. Zukowski, A. Zeilinger, M. Horne, and A. Ekert, Phys. Rev. Lett. \textbf{71}, 4287 (1993)

\bibitem{pan-bouwmeester-weinfurter-zeilinger}
J.-W. Pan, D. Bouwmeester, H. Weinfurter, and A. Zeilinger, Phys. Rev. Lett. \textbf{80}, 3891 (1998)

\bibitem{HanburyBrownTwiss}
R. Hanbury Brown and R. Q. Twiss, Proc. of the Royal Society of London, A \textbf{242}, 300 (1957)

\bibitem{henny-etal99}
M. Henny, S. Oberholzer, C. Strunk, T. Heinzel, K. Ensslin, M. Holland, and C. Sch\"onenberger, Science \textbf{284}, 296 (1999)

\bibitem{bromberg-etal2010}
Y. Bromberg, Y. Lahini, E. Small, and Y. Silberberg, Nature Photonics \textbf{4}, 721 (2010)

\bibitem{bertlmann-hiesmayr2006}
R. A. Bertlmann and B. C. Hiesmayr, Quantum Information Processing \textbf{5}, 421 (2006)

\bibitem{reeh-schlieder}
H. Reeh and S. Schlieder, Nuovo Cim. \textbf{22}, 1051 (1961)

\bibitem{haag-book}
R. Haag, \emph{Local quantum physics: Fields, particles, algebras}, Springer 1992

\bibitem{narnhofer-PLA2003}
H. Narnhofer, Phys. Lett. A \textbf{310}, 423 (2003)

\bibitem{fuentes-mann}
I. Fuentes-Schuller and R. B. Mann, Phys. Rev. Lett. \textbf{95}, 120404 (2005)

\bibitem{alsing-fuentes-etal}
P. M. Alsing, I. Fuentes-Schuller, R. B. Mann, and T. E. Tessier, Phys. Rev. A \textbf{74}, 032326 (2006)

\bibitem{friis-b-h-h}
N. Friis, R. A. Bertlmann, M. Huber, and B.C. Hiesmayr, Phys. Rev. A \textbf{81}, 042114 (2010)

\bibitem{friis-k-m-b}
N. Friis, P. K\"ohler, E. Mart$\acute{\i}$n-Mart$\acute{\i}$nez, and R. A. Bertlmann, arXiv:1107.3235 [quant-ph]

\bibitem{smith-mann}
A. Smith and R. B. Mann, arXiv:1107.4633 [quant-ph]

\bibitem{narnhofer-ClassQuantGrav2011}
H. Narnhofer, Class. Quantum Grav. \textbf{28}, 145016 (2011)

\bibitem{narnhofer-tobepub}
H. Narnhofer, to be published

\bibitem{laughlin1981}
R. B. Laughlin, Phys. Rev. B \textbf{23}, 5632 (1981)

\bibitem{narnhofer-RepMathPhys2004}
H. Narnhofer, Rep. Math. Phys. \textbf{53}, 93 (2004)

\bibitem{zoller_ESI-lecture}
P. Zoller, \emph{Quantum Computing and Quantum Simulation with Quantum Optical Systems}, Lecture given within the Erwin Schr\"odinger Symposium at The Erwin Schr\"odinger International Institute for Mathematical Physics, January 13 - 15, 2011

\end{thebibliography}
\end{document}